\documentclass[twocolumn]{aastex631}

\usepackage{amsmath}
\usepackage{comment}

\usepackage[encapsulated]{CJK}
\usepackage{hyperref}
\usepackage{url}

\providecommand{\apjl}{ApJL}

\providecommand{\apjs}{ApJS}

\providecommand{\aap}{A\&A}

\begin{document}
\begin{CJK}{UTF8}{bsmi}
\title{A MUltiwavelength Study of ELAN Environments (AMUSE$^2$): \\
The Impact of Dense Environment on Massive Dusty Star-Forming Galaxies at Cosmic Noon
}

\correspondingauthor{Yu-Jan Wang, Chian-Chou Chen}
\email{yujanwang@asiaa.sinica.edu.tw, ccchen@asiaa.sinica.edu.tw}

\author{Yu-Jan Wang(王禹然)}
\affiliation{Graduate Institute of Astrophysics, National Taiwan University, Taipei, Taiwan}
\affiliation{Academia Sinica Institute of Astronomy and Astrophysics (ASIAA), No. 1, Section 4, Roosevelt Road, Taipei 106216, Taiwan}

\author{Chian-Chou Chen(陳建州)}
\affiliation{Academia Sinica Institute of Astronomy and Astrophysics (ASIAA), No. 1, Section 4, Roosevelt Road, Taipei 106216, Taiwan}

\author{Fabrizio Arrigoni Battaia}
\affiliation{Max-Planck-Institut f\"ur Astrophysik, Karl-Schwarzschild-Str. 1, D-85748 Garching bei M\"unchen, Germany}

\author{Roberto Decarli}
\affiliation{INAF − Osservatorio di Astrofisica e Scienza dello Spazio di Bologna, via Gobetti 93/3, I-40129 Bologna, Italy}

\author{Helmut Dannerbauer}
\affiliation{Instituto de Astrof{\'i}sica de Canarias (IAC), 38205 La Laguna, Tenerife, Spain}

\author{Po-Feng Wu(吳柏鋒)}
\affiliation{Graduate Institute of Astrophysics, National Taiwan University, Taipei, Taiwan}

\begin{abstract}
To understand how massive galaxies are influenced by their surroundings, we present new ALMA and NOEMA observations as part of A MUltiwavelength Study of ELAN Environments (AMUSE$^2$). These observations target submillimeter sources discovered in single-dish surveys around nine quasars hosting Ly$\alpha$ nebulae at $z=2\sim3$, including two Enormous Ly$\alpha$ nebulae (ELANe). Through detection of mid-$J$ CO lines, we confirm physical associations of 15 SMGs, which are located outside the expected virial radii of the central dark-matter halos hosting the quasars. We find $73^{+29}_{-21}\%$ of SMGs have line profiles better described by double Gaussian models, with a median peak-to-peak separation of 350 $\pm$ 25 km/s, suggesting rotating disks or interacting pairs. Modified blackbody fits of the far-infrared photometry yield a median $\beta$ of 2.0 $\pm$ 0.2 and $T_{dust}$ of 34 $\pm$ 3 K. Overall, SMGs outside quasar halos share similar physical properties with those in the field, but combining data from other studies reveals depleted gas fractions within quasar halos. This suggests that dense environments significantly impact massive star-forming galaxies only within halo scales at cosmic noon. Additionally, spatial analyses of 15 SMGs indicate they trace large-scale structures, possibly filamentary or elongated pancake-like, with a scale height of 2-5\,cMpc. Our measured distributions and densities of star-formation rates align with models, though likely represent lower limits.
\end{abstract}

\section{Introduction} \label{sec:intro}

It has long been demonstrated that galaxy clusters are dominated by massive elliptical galaxies in the local universe. This suggests that the environment of clusters has a significant impact on the transition of star-forming galaxies to quiescent galaxies. 
According to \citealp{1980ApJ...236..351D}, their analyses of morphology in 55 rich clusters in the local Universe indicated a well-constrained relationship between density and the morphological properties of galaxies. Additionally, \citealp{2010ApJ...721..193P}, who obtained a strong relation between the quenching rate and the local density of galaxies, also point out that several mechanisms may quench galaxies from early epochs to what they look like today, revealing that star formation activity in most galaxies could be influenced by internal factors such as their gas reservoir, as well as external processes including galaxy interactions. 
However, galaxies within clusters in the local universe are already quenched. Thus, to understand how galaxies evolve and quench star formation within galaxy clusters, it is essential to comprehend the progenitors of the dense environments in the local universe, or the proto-clusters \citep{Chiang:2013aa,Overzier:2016aa}, and more fundamentally, how the environment impacts galaxy formation in the early Universe. 

Looking back into the higher redshift epoch at $z=2-4$, known as ``cosmic noon'', it is still unclear whether and how the environment connects to galaxies and their properties. For example, from the perspective of gas, some studies have found that the gas properties of galaxies change depending on the environment \citep{2018ApJ...856..118H,Alberts2022}. In particular,  
 recently \citet{Hughes:2025aa} found an increase in gas density and excitation temperature based on a [C{\sc i}] survey of 25 dusty star-forming galaxies in the core of the SPT2349-56 protocluster at z = 4.3. However, on the other hand, some studies did not find significant evidence that the environment impacts gas properties \citep{2017A&A...608A..48D,2017ApJ...842...55L}, or that the impact could be mass dependent \citep{Tadaki:2019aa}.
 Therefore, it can be seen that the results of observational studies have thus far been inconclusive. This can potentially be attributed to a variety of factors, including the sample selection process, the methodological framework employed, and the quality of the data obtained \citep{Alberts:2022bb}.



On the other hand, simulations of structure formation in the Universe have predicted that galaxies are embedded in the ``cosmic web'' (\citealp{2014MNRAS.441.2923C, 2024A&A...684A..63G}). Galaxies evolve via mergers and matter accretion from large-scale filaments that build up the web. The detection of cosmic web filaments has been reported both directly from emissions of gas or galaxies \citep{Erb:2011aa, Cantalupo:2014aa, Umehata:2019aa, FAB:2019aa,2021A&A...652A..11J, Johnson:2022aa, Emonts:2023aa, Sun:2024aa, Herwig:2024, Tornotti:2024}, or indirectly via tomography of the Ly$\alpha$ forest \citep{Lee:2014aa, Newman:2020aa, Horowitz:2022aa}. Thanks to surveys of ALMA and NOEMA, dusty star-forming galaxies (DSFGs), or submillimeter galaxies (SMGs)\footnote{ In this paper we use SMGs to specifically refer to the DSFGs that are selected at 850/870\,$\mu$m.}, have also been found to trace large-scale filament-like structures. 
In fact, recent studies based on SCUBA-2 surveys around quasars at $z\sim2-3$ have found filament- or pancake-like structures \citep{FAB2023, 2024arXiv240616637W}, and the width of these structures is $\sim$3-4 cMpc, consistent with model predictions of thick filaments traced by massive halos. DSFGs are massive infrared luminous galaxies located at cosmic noon and beyond \citep{Chapman:2005aa,Simpson:2014aa,Chen:2022aa,Cox:2023aa,Long:2024aa,Berta:2025aa}. They are hosted by halos with masses similar to those hosting quasars \citep{Hickox:2012aa,Chen:2016aa,Wilkinson:2017aa,Stach2021,Lim:2021aa}, suggesting co-evolution between these two massive populations that eventually leads to the formation of local ellipticals or clusters \citep{2020MNRAS.494.3828D,Amvrosiadis:2025aa}. Systematic surveys of DSFGs around quasars could help advance our understanding of their co-evolution.

Finally, several recent cosmological hydrodynamical simulations have attempted to replicate the total SFRs in proto-clusters, but emphasize the presence of tensions between observations and simulations. For example, \citet{2020A&A...642A..37B} studied SFRs in simulations of galaxy clusters at $z\sim0-4$, showing that they underpredict by a factor $\geq$ 4 the star formation both at z $\sim$ 2 and z $\sim$ 4. Subsequently, \citet{2022MNRAS.509.4037Y} conducted simulations of proto-cluster regions (FOREVER22), predicting the cumulative SFR at $z=3$ and the evolution of the SFR density up to $z\sim10$. However, these simulation models yield relatively small values compared to the observed total SFRs in proto-clusters. One possible explanation of such tensions could be that the existing sub-grid physics used for simulating star formation in these models may not be suitable for dusty starbursts (\citealp{2023ApJ...950..191R}), which are the main tracers in observational studies. Indeed, the observed measurements might be biased towards proto-clusters with a high prevalence of dusty starbursts, influenced by selection biases. In addition, it could be that the volumes within which the calculations are performed are not consistently defined. Therefore, a comprehensive survey of star-forming populations would help us better understand the current difference between models and observations.

Our study targets quasars with known extended Ly$\alpha$ emission (\citealt{FAB:2019}), where the most extreme are called enormous Ly$\alpha$ nebulae (ELANe; \citealt{2014Natur.506...63C,2015Sci...348..779H,2017ApJ...837...71C,2018MNRAS.473.3907A}). Fields centered at these quasars at $z\sim2-4$ are found to be good candidates for proto-cluster fields \citep{Umehata:2019aa,Nowotka2022,FAB2023,2024A&A...684A.119P}.
By exploiting 850\,$\mu$m JCMT/SCUBA-2 observations, \citet{Nowotka2022} and \citet{FAB2023} found that submillimeter sources around these quasars are overabundant with respect to blank fields by an average factor of $\sim3.5$. 
However, these surveys at 850\,$\mu m$ could only provide us with the distribution in 2D projected space, and spectroscopic follow-up observations are needed to confirm if these submillimeter sources are physically associated with the central quasars. In this work, we thus search for CO emission lines of a subsample of the brightest SMGs in these fields using ALMA and NOEMA observations. The ultimate goal is to determine their redshifts (hence association with the quasars) and quantify their ISM properties (e.g., molecular gas and dust content) to verify the presence of any environmental impact on the evolution of such galaxies.

This paper is structured as follows. In Section \ref{sec:Observation}, we describe observations and data reduction of the sample used in this paper. In Section \ref{sec:Analyses and Results}, the methods to extract CO lines and to obtain the ISM properties are presented. In Section \ref{sec:Discussion}, we present the results in three main parts: (i) Gas and dust properties: we compare the ISM properties between the SMGs surrounding the quasar hosting Ly$\alpha$ nebulae and those in the field, studying the environmental impact on the properties of galaxies. (ii) Spatial distribution: we analyze the spatial distributions of SMGs around quasar fields to investigate the large-scale filamentary structures. (iii) Cumulative star formation rate: We compare the star formation rate densities of our samples to other overdense regions and simulation results. Finally, a summary is presented in Section \ref{sec:Conclusion}. Throughout this paper, we assume the Planck cosmology: $H_0$ = 67.8 $km s^{-1} Mpc^{-1}$ , $\Omega_MΛ$ = 0.307, and $\Omega_\Lambda$ = 0.69 \citep{2014A&A...571A..16P}. 

\section{Observations and data} \label{sec:Observation}

\subsection{Interferometer data} \label{subsec:interferodata}

\subsubsection{ALMA} \label{subsec:alma}
The ALMA observations for this work were carried out from November 2019 to January 2020 (PID: 2019.1.01514.S
). The target submillimeter sources were selected as significant detections (SNR$\ge4.5$) in the original SCUBA-2 survey that targeted $z\sim2-3$ ELAN and quasar fields (\citealt{Nowotka2022,FAB2023}). In total we target 91 objects around eight quasar fields with known Ly$\alpha$ extended emission. These eight quasar fields were the first to be observed in our SCUBA-2 survey. 
Data were taken with 36 to 46 12-meter antennas, and the baselines ranged from 15.0 to 783.5 meters. The mean precipitation water vapor (PWV) varied from 1.1 to 6.3\,mm depending on the night. For each execution block (EB), the total on-source time (ToS) is approximately 40 to 60 minutes, totaling about 6 hours, as reported in the QA2 report. 
The observational setup used four spectral windows, one window tuned to the redshift of the central quasar to search for the CO(4-3) emission line of the SMGs, and the other three windows capturing their continuum. Since the central quasars are located at $z=2-3$, these spectral windows are in band 3 (for 6 fields) and band 4 (for 2 fields).
The radius of the primary beam for band 3 and band 4 with the 12m array is 40$\arcsec$ and 32$\arcsec$, respectively. The details of each field and their corresponding calibrators are shown in \autoref{tab:tab1}.

\begin{table*}[!ht]
    \centering
    \caption{Information about ALMA and NOEMA observations}
    \setlength{\tabcolsep}{2mm}{
    \begin{tabular}{lcccc}
    \toprule
        Field name & $z_{\rm QSO}$ & number of source & Calibrator &   \\
           & & detected$^a$/observed  &Flux/Bandpass &  Phase  \\\hline
        (ALMA)  &  &\\
        Q2355 & 3.385 & 3/9 & J0006-0623 & J2337-0230 \\
        SDSSJ0819 & 3.197 & 7/16 & J0854+2006 \& J0750+1231 & J0818+0517 \\ 
        SDSSJ1020 & 3.164 &  3/14 & J1058+0133 & J1015+1227 \\ 
        SDSSJ1025 & 3.227 &  5/11 & J1058+0133 & J1016+0513 \\ 
        SDSSJ1209 & 3.117 &  4/14 & J1256-0547 & J1206+0529 \& J1222+0413 \\ 
        SDSSJ1342 & 3.062 &  4/7 & J1256-0547 &  J1357+1919 \\
        MAMMOTH$^b$ & 2.317 & 0/15  & J1550+0527 & J1419+3821 \\
        UM287$^b$ & 2.267 & 0/5  &  J0006-0623 & J0049+0237\\ \hline
        (NOEMA) &  &  & \\
        Jackpot & 2.0412 & 1/10 & 3C84 \& 0420-014 \& LKHA101  & 0821+394 \&
0923+392\\ \hline
    \end{tabular}}
    \tablecomments{$^{a}$ the number of detected sources include both primary and supplementary set. $^b$observation in band 4}
    \label{tab:tab1} 
\end{table*}

For data reduction, we calibrate the raw visibilities by running the corresponding \textsc{scriptForPI.py} scripts under \textsc{CASA5.6.1}. This means that the calibrations performed by the QA2 team members are adopted. During the imaging process, we first perform continuum
subtraction. Specifically, we exclude channels that include significant line emissions.
The continuum channels are then used for both continuum generation and subtraction. It should be noted that although we designate all band 3 as continuum at 3\,mm, these continua may have slight offsets in wavelength from 3\,mm depending on the tuning of different observations. For band 4, the wavelength of the continuum was designated as 2\,mm.

Next, we apply \textsc{Tclean} to perform an inverse Fourier transformation on both the continuum and continuum-subtracted visibilities. The final images are transformed into 240$\times$240 pixels in R.A. and DEC. directions, respectively, with a pixel size of 0$\farcs$43. We apply natural weighting for the baselines. Following the imaging method described in \citealp{2022ApJ...929..159C}, we use an automasking approach in \textsc{Tclean} by setting “auto-multithresh” as the usemask parameter to increase the efficiency of CLEAN (noisethreshold = 4; lownoisethreshold = 2.5). Finally, the cubes within the masked regions are cleaned down to 2\,$\sigma$ level. During these steps, for each source we output two cleaned data cubes, each of which represents the combined adjacent spectral windows 
in the upper sideband (USB), covering the redshifted CO lines of the central quasar with a velocity range of $\sim$ -2000 to 8000 km/s, and the lower sideband (LSB). Both sidebands contain 256 channels.

Because the resolution of each channel differs depending on frequency, we apply the \textsc{CASA} routine \textsc{Imsmooth} to the reduced data cubes, allowing each cube to have a common resolution. This common resolution is typically the largest beam size of the pre-smoothed cube. Subsequently, we derive the cube with spatial resolution (beam size) in major axis
ranged from $\sim$2$\farcs$0 - 4$\farcs$3 for USB, and $\sim$2$\farcs$2 - 4$\farcs$7 for LSB. The spatial resolution of the continuum is the average value between the USB and LSB. The final cleaned continuum image has a sensitivity (1-$\sigma$ noise level) of $\sim40\,\mu$Jy\,beam$^{-1}$, and the mean sensitivity for the cleaned image cube is $\sim0.4$\,mJy\,beam$^{-1}$ per channel of 15 MHz ($\sim$40\,km\,s$^{-1}$). 

\subsubsection{NOEMA} \label{subsec:noema}

NOEMA observations were secured on 
June 13, 16 and 24 2019 using the compact (9D) configuration array, with 10 antennas, as part of the 
S19CK program (PI: F. Arrigoni Battaia). The weather conditions were good, with modest wind ($<10$ km\,s$^{-1}$) and precipitable water vapour (1--3 mm). The targets were observed in about 3 hours per source in track-sharing mode. The radio sources 3C84, 0420-014, and LKHA101 were used for flux and bandpass calibration; 0821+394 and 0923+392 for phase and amplitude calibration. We used the software \textsc{Clic} from the \textsc{Gildas} suite to reduce and calibrate the data, resample the observations in 20\,MHz--wide channels, and extract the calibrated visibilities. We then used \textsc{Mapping} to perform the Fourier transform.
We produce cubes with a pixel size of 0$\farcs$6 and a spatial resolution (beam size) in major axis
ranging from $\sim$3$\farcs$8 for USB to $\sim$4$\farcs$4 for LSB. The spatial resolution of the continuum is the average value between the USB and LSB. The final cleaned continuum image has a sensitivity (1-$\sigma$ noise level) of $\sim40\,\mu$Jy\,beam$^{-1}$, and the mean sensitivity for the cleaned image cube is $\sim0.4$\,mJy\,beam$^{-1}$ per channel of 20 MHz ($\sim$50\,km\,s$^{-1}$).

\subsection{Multi-wavelength data} \label{subsec:multilambda}

\subsubsection{JCMT/SCUBA-2} \label{subsec:SCUBA-2}
To aid our analyses, we make use of the SCUBA-2 archival data at both 450 and 850\,$\mu$m. The details of these data can be found in \citet{FAB2023}. We briefly summarize them in the following. The observations were taken using the Daisy scanning pattern, in each field covering approximately a circular footprint with 14$'$ in diameter, centered at our sample quasars. We perform data reduction following \citet{2024ApJ...971..117G} and the reduced images have their central noise levels (1\,$\sigma$) reaching the range 3.9 - 20.5 mJy\,beam$^{-1}$ and 0.5 - 1.1 mJy\,beam$^{-1}$ at 450 and 850\,$\mu$m, respectively. Our results are consistent with those reported in the original survey \citep{FAB2023}.

\subsubsection{Herschel} \label{subsec:herschel}
To increase the coverage of the spectral energy distribution (SED) for our analyses, we found Herschel archival data taken on some of our quasar fields. In particular, we utilize the SPIRE data for Q2355 (HerMES Large Mode Survey; PI: Viero, M.; GT2\_mviero\_1) and SDSSJ0819 (PI: Weedman, D.; OT2\_dweedman\_2). We retrieved the data from the ESA Herschel Science Archive, and we simply used the level-2 map reduced by the pipeline from the archive at 250, 350, and 500\,$\mu$m. In the level-2 map, the mean sensitivities are 17.9/15.0/18.9\,mJy\,beam$^{-1}$ and 13.0/11.0/14.0\,mJy\,beam$^{-1}$ at 250/350/500\,$\mu m$, for Q2355 and SDSSJ0819, respectively.

\section{Analyses and Results} \label{sec:Analyses and Results}
\subsection{CO Line Emission} \label{subsec:Line}

\begin{figure*}[h]
\centering
\includegraphics[width=0.95\textwidth]{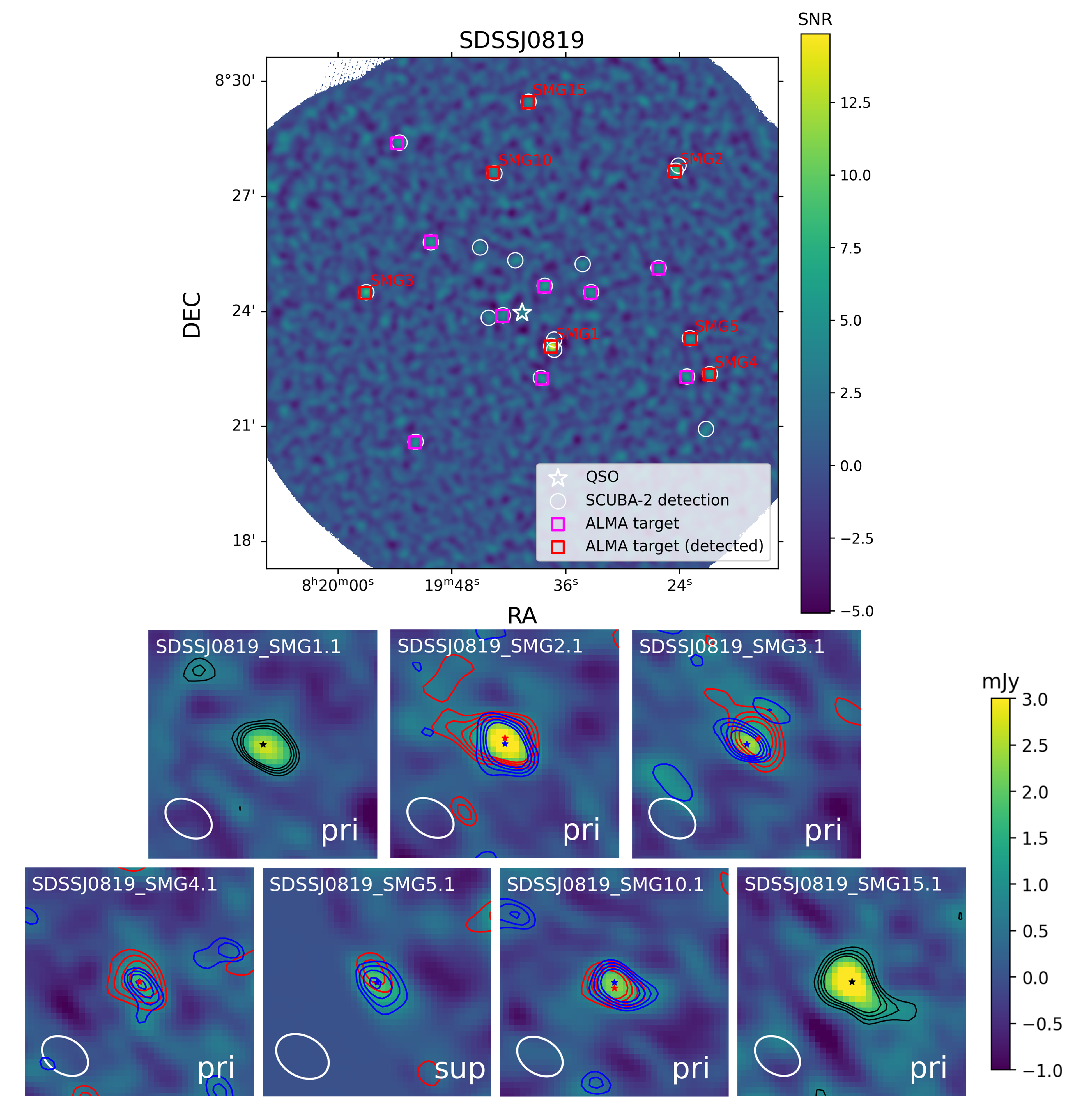}
\caption{An example of targeted quasar field SDSSJ0819 and its ALMA detections. \textbf{Upper panel:} SCUBA-2 850\,$\mu$m map where 
white circles indicate the $\geq 4$\,$\sigma$ detections at 850\,$\mu$m reported by \citet{FAB2023}. Magenta squares represent the ALMA targets presented in this work. Targets with CO detections are labeled in red with their corresponding names. \textbf{Lower panels:} Integrated CO emission maps 
integrated over channels within the FWHM of the detected emission line. 
White circles illustrate the beam size and shape of the corresponding targets. Images featuring only black contours indicate that the lines are better described by single-Gaussian (SG) models (Section~\ref{subsec:Line}), with the contours representing 3-6\,$\sigma$ of the stacked maps. Conversely, images with blue and red contours represent emission lines that are better fitted by double-Gaussian (DG) models. Blue and red contours denote 3-6 $\sigma$ of the stacked map based on each of the two components of the DG models, respectively.
Figures for the rest of the detections are shown in \autoref{apx:a}-\ref{apx:b}.
\label{fig:fig.1}}
\end{figure*}

\begin{figure*}[h]
\centering
\includegraphics[width=\textwidth]{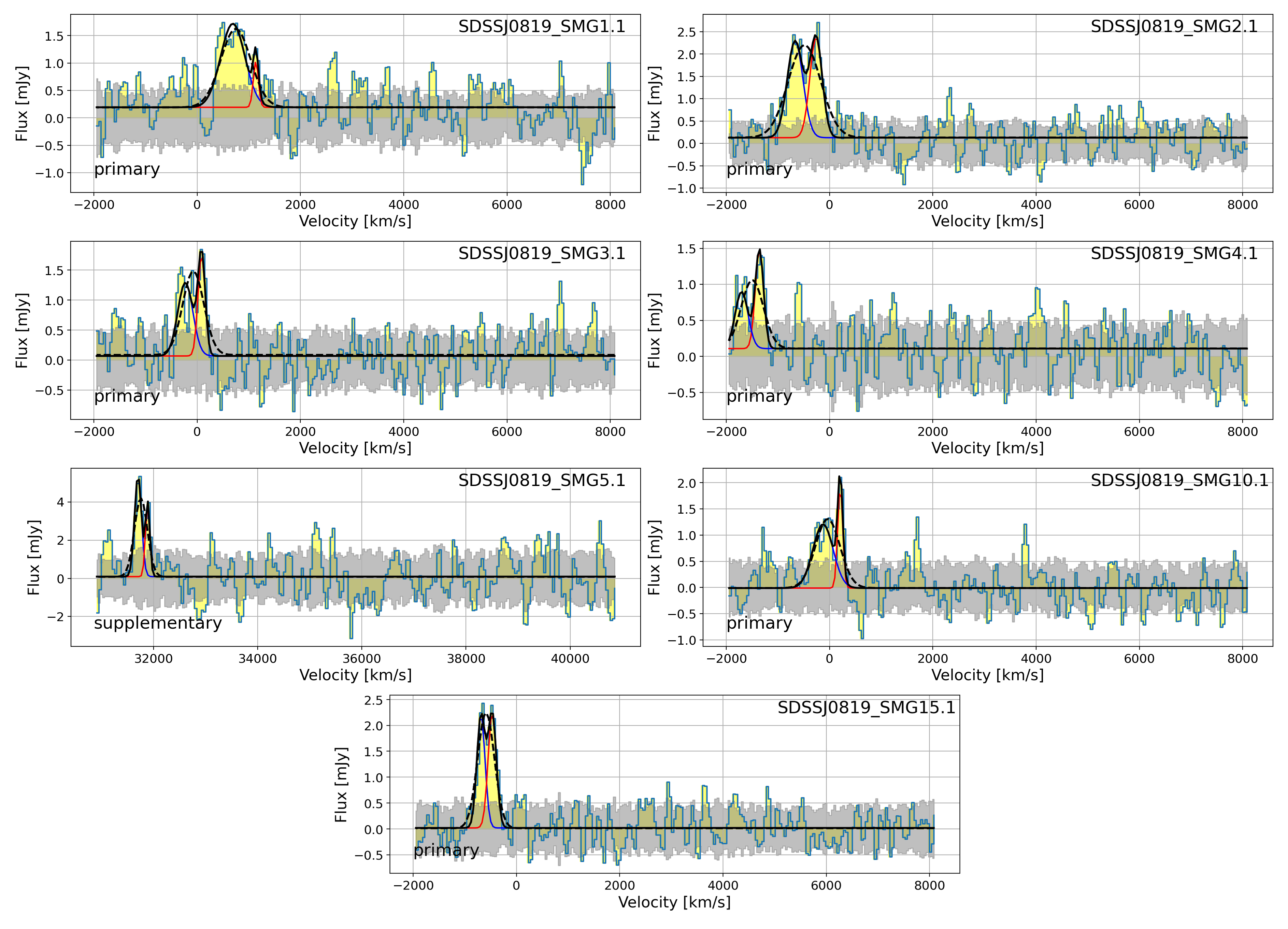}
\caption{Spectra of the detected sources and their fitting results of SDSSJ0819.
The gray region depicts the noise for each channel. Fitting results for both the single-Gaussian (SG) and double-Gaussian (DG) models are shown on the figure: the black dotted lines represent the best-fit SG models, while the black solid line represents those for DG, with the two components displayed separately as blue and red solid lines.
Figures for the rest of the detections are shown in \autoref{apx:c}.
\label{fig:fig.1b}}
\end{figure*}

\subsubsection{Line extraction from spectral datacubes} \label{subsubsec:311}

To search for line emitting sources 
in the reduced data cubes, we utilized the publicly available code \textsc{LineSeeker}, previously employed for searching line emission in other ALMA spectral cubes (\citealp{2017A&A...597A..41G}). We consider line detections to be signal-to-noise ratios (SNRs) over 7.0 - 8.0 (as detailed in Section \ref{subsubsec:313}), exceeding the maximum SNRs of the negative signals. Similar SNR thresholds were adopted in previous line searches of other ALMA cubes (\citealp{2019ApJ...882..136A, 2019ApJ...882..139G}). After removing spikes detections that have FWHM values below the minimum ($\sim$180 km/s) of the reported lines for SMGs in the literature  \citep{2021MNRAS.501.3926B}, a total of 27 significant line detections were obtained from the ALMA cubes (21 in USB, 6 in LSB), and 1 significant line detection from the NOEMA cube. 
\par 
We classified these 28 line detections into two types: primary and supplementary. Primary sources are those that are more likely to be genuine line emissions based on their continuum detections
(see Section \ref{subsec:32}). In addition to having continuum detection, the primary sources are also those that are likely located within the same large-scale structures of the central quasars. According to previous studies in protocluster environment (\citealp{2021A&A...652A..11J}; \citealp{2024arXiv240616637W}), the standard deviation of relative velocity between central QSO and the member galaxies is $\sim$ 2500 km/s. Simply considering the normal distribution, it reach 3-$\sigma$ level when velocity reach $\sim\pm$7000 km/s. This velocity range corresponds to $\sim\pm 90$\,cMpc in distance for the Hubble flow, which is within the filament lengths predicted in recent simulations \citep{Galarraga-Espinosa:2024aa}. Therefore, the relative velocity of primary samples is restricted to this range, 
ensuring their membership within the large-scale structure of the targeted proto-clusters. In short summary, primary sources are those that are detected in continuum and have velocity offsets within $\pm$7000\,km/s with respect to their corresponding quasars.

Conversely, supplementary sources lack other significant evidence supporting their fidelity. Detections are also classified as supplementary if their relative velocity to the corresponding quasars exceeds the range of $\sim\pm$7000 km/s, resulting in them being likely serendipitous detections along the line of sight even though they may be genuine. 
To understand the nature of these serendipitous detections, we conducted an analysis for the number density of reliable supplementary sources, and found agreement with the reported CO luminosity functions in the field (see Appendix \ref{apx:d}).   
Therefore, we focus our analyses on the primary set of 15 sources, and we provide information for both primary and supplementary samples in \autoref{tab:tab3}. 
\par
We extract spectral lines using apertures of the synthesized beams at the locations reported by \textsc{LineSeeker}. We make corrections to total line intensities following a curve-of-growth analysis  (Section \ref{subsubsec:312}). To determine if the spectra are better fit by a single Gaussian or double Gaussian model, we employed the Akaike information criterion (AICc) (as detailed in \citealp{2022ApJ...929..159C}).
The best-fit model results are shown in Table \ref{tab:tab3}. We find that $73^{+29}_{-21}\%$ (considering Poisson error) of our primary samples are better described by double Gaussian models, consistent with previous findings for the field SMGs \citep{2021MNRAS.501.3926B,2022ApJ...929..159C}. Additionally, we derive the median separation of the two Gaussian peaks to be $350\,\pm\,25\ {\rm km\  s^{-1}}$, again reaching a comparable level to the value reported by \citet{2022ApJ...929..159C}, suggesting similar dynamical properties between SMGs in dense regions and those in the field.

\par
To derive the locations of CO emission lines, we stacked the channels that fall within the FWHM determined by the best model to generate the line emission maps (\autoref{fig:fig.1} and \autoref{apx:b}). We then identified the locations of peak flux density in these maps and the results are reported in \autoref{tab:tab3}. For sources that are better fit by the double Gaussian model, we also separately stacked the channels corresponding to each component, indicated by the red and blue stars in the lower panels of \autoref{fig:fig.1}. 
Although there are offsets between the redshifted and blueshifted components, the offsets are smaller than the beam size. Therefore, we do not find evidence of significant offsets between the two components at the given spatial resolution, but there appears to be a hint of disk rotation in these galaxies.
\par
With the chosen best-fit model for the CO emission line, we then derived line properties by integrating the model. For single Gaussian models, redshifts are determined based on the frequency of the flux peak channel, and the FWHM is 2.355 times the standard deviation. For the double Gaussian model, we determined the redshifts using the mean frequency defined as
\begin{displaymath}
\bar{\nu} = \frac{\int \nu I_{\nu} \,d\nu}{\int I_{\nu} \,d\nu},
\end{displaymath} and the FWHM was determined as 2.355 times the dispersion defined as  
\begin{displaymath}
\bar{\sigma} =\sqrt{ \frac{\int (\nu-\bar{\nu})^2 I_{\nu} \,d\nu}{\int I_{\nu} \,d\nu}}
\end{displaymath}
where $I_{\nu}$ represents the flux density at the corresponding frequency. The results can be found in Table \ref{tab:tab3}.
\par
Given the redshifts of the central quasars, we identify these line detections to be most likely CO(4-3) from ALMA and CO(3-2) from NOEMA. 
Based on the redshift-dependent probability distributions of field SMGs (e.g., \citealt{2016ApJ...820...82C}), we consider the conversion between redshift range and frequency of our spectral coverage in different CO transitions, and finding that the probability of these lines being other CO transitions in chance projection is $\sim$10\% (e.g., $\sim$8\% for ALMA detections being CO(3-2)). Finally, with the intensity of the CO emission lines available, we convert them to the CO luminosity following the method outlined by \citet{2013ARA&A..51..105C}, defined as  
\begin{displaymath}
L'_{CO} = \frac{3.257\times{10}^7}{1+z}\frac{I_{CO}D_L^2}{\nu_0^2}\  [{\rm K\,km\,s^{-1}\,{pc}^{2}}]
\end{displaymath}
where $z$ represents the redshift, $I_{\rm CO}$ is the intensity derived in the previous paragraph with units in Jy\,km\,$\rm s^{-1}$, $D_{\rm L}$ is the luminosity distance in Mpc, and $\nu_0$ is the mean frequency in GHz. The results can also be found in Table \ref{tab:tab3}. Additionally, for comparison with other samples, we convert $L'_{\rm CO(3-2)}$ 
and $L'_{\rm CO(4-3)}$ 
into $L'_{\rm CO(1-0)}$ using the relationship described by \cite{2021MNRAS.501.3926B}: $L'_{\rm CO(1-0)}=L'_{\rm CO}/r_{j1}$, with $r_{31}= 0.63\pm0.12$ 
; $r_{41}= 0.34\pm0.04$ 
. We then determine that the median value of $L'_{\rm CO(4-3)}$ for the primary set is $(3.9\pm0.5)\times 10^{10}$\ K\ km\ s$^{-1}$\ pc$^{2}$, with errors computed via bootstrapping.

\begin{table*}[!ht]
    \centering
    \caption{Locations and properties of CO emitters, split by their corresponding categories. The supplementary sources are further split based on the origin of their data.}
    \setlength{\tabcolsep}{1.7mm}{
    \begin{tabular}{lccccccccccccccccc}
    \toprule
        Primary \\
        ID(ALMA) & RA & Dec & S/N & $z_{spec}$ & model & FWHM & $I_{\rm CO}$ & $L'_{\rm CO(4-3)}$ \\
         ~&~&~&~&~&~&[km/s]&[Jykm/s]&[$10^{10} K km s^{-1} pc^{2}$] \\ \hline
        Q2355\_SMG8.1 & 23:57:51.48 & 01:23:21.6 & 11.6 & 3.4581 ± 0.0003 & DG & 412 ± 24 & 1.4 ± 0.2 & 4.3 ± 0.6 \\ 
        Q2355\_SMG9.1 & 23:58:07.90 & 01:29:14.9 & 10.7 & 3.4119 ± 0.0013 & DG & 980 ± 90 & 1.4 ± 0.3 & 4.4 ± 1.0 \\ 
        SDSSJ0819\_SMG1.1 & 08:19:37.23 & 08:23:06.1 & 13.2 & 3.2075 ± 0.0006 & SG &685 ± 96 & 2.1 ± 0.5 & 6.0 ± 1.4  \\ 
        SDSSJ0819\_SMG2.1 & 08:19:24.05 & 08:27:43.4 & 18.1 & 3.1902 ± 0.0005 & DG &576 ± 42 & 3.0 ± 0.5 & 8.5 ± 1.3  \\ 
        SDSSJ0819\_SMG3.1 & 08:19:56.90 & 08:24:31.0 & 10.4 & 3.1956 ± 0.0007 & DG &460 ± 67 & 1.5 ± 0.4 & 4.2 ± 1.2 \\ 
        SDSSJ0819\_SMG4.1 & 08:19:20.56 & 08:22:24.8 & 8.7 & 3.1758 ± 0.0008 & DG &475 ± 72 & 1.0 ± 0.3 & 3.0 ± 0.9  \\ 
        SDSSJ0819\_SMG10.1 & 08:19:43.54 & 08:27:39.3 & 10.4 & 3.1966 ± 0.0006 & DG &529 ± 75 & 1.7 ± 0.4 & 4.8 ± 1.1  \\ 
        SDSSJ0819\_SMG15.1 & 08:19:39.82 & 08:29:30.2 & 14.3 & 3.1888 ± 0.0007 & SG &384 ± 40 & 1.8 ± 0.3 & 5.2 ± 0.9  \\ 
        SDSSJ1020\_SMG3.1 & 10:20:28.94 & 10:39:33.7 & 11.6 & 3.2245 ± 0.0008 & DG &575 ± 81 & 2.0 ± 0.5 & 5.7 ± 1.3  \\ 
        SDSSJ1025\_SMG1.1 & 10:25:17.85 & 04:52:33.4 & 10.8 & 3.2368 ± 0.0005 & SG &550 ± 87 & 1.2 ± 0.3 & 3.5 ± 0.8  \\
        SDSSJ1209\_SMG2.1 & 12:09:20.52 & 11:39:05.7 & 9.8 & 3.1243 ± 0.0005 & DG &336 ± 51 & 1.2 ± 0.3 & 3.2 ± 0.8  \\ 
        SDSSJ1209\_SMG4.1 & 12:09:29.38 & 11:36:05.7 & 14.0 & 3.1515 ± 0.0003 & SG &571 ± 58 & 2.3 ± 0.4 & 6.4 ± 1.0  \\ 
        SDSSJ1342\_SMG1.1 & 13:42:29.50 & 17:01:40.6 & 15.5 & 3.0928 ± 0.0004 & DG &487 ± 31 & 1.5 ± 0.2 & 4.0 ± 0.6  \\ 
        SDSSJ1342\_SMG5.1 & 13:42:46.45 & 17:05:03.8 & 12.5 & 3.0294 ± 0.0007 & DG &710 ± 64 & 1.7 ± 0.3 & 4.5 ± 0.8  \\ \hline
        Primary \\
        ID(NOEMA) & RA & Dec & S/N & $z_{spec}$  &model& FWHM & $I_{\rm CO}$ & $L'_{\rm CO(3-2)}$ \\
        ~&~&~&~&~&~&[km/s]&[Jykm/s]&[$10^{10} K km s^{-1} pc^{2}$] \\ \hline
        Jackpot\_SMG5.1 & 08:41:46.54 & 39:20:05.2 & 9.2 & 2.1051 ± 0.0022 & DG & 781 ± 155 & 2.4 ± 1.0 & 5.532 ± 2.279 \\  
        \hline\hline
        Supplementary  \\
        ID(ALMA, USB) & RA & Dec & S/N & $z_{spec}$  &model & FWHM & $I_{\rm CO}$ & $L'_{\rm CO(4-3)}\ ^a$ \\
         ~&~&~&~&~&~&[km/s]&[Jykm/s]&[$10^{10} K km s^{-1} pc^{2}$] \\ \hline
        SDSSJ1020\_SMG10.1 & 10:20:10.68 & 10:38:42.7 & 8.7 & 3.2376 ± 0.0003 & DG &389 ± 29 & 2.3 ± 0.5 & 6.8 ± 1.5 \\ 
        SDSSJ1025\_SMG4.1 & 10:24:58.17 & 04:51:26.8 & 8.8 & 3.2748 ± 0.0004 & DG &301 ± 21 & 2.3 ± 0.8 & 6.9 ± 2.4  \\ 
        SDSSJ1025\_SMG4.2 & 10:24:58.20 & 04:50:43.8 & 8.6 & 3.2699 ± 0.0006 & DG &392 ± 73 & 1.0 ± 0.3 & 3.0 ± 0.9  \\ 
        SDSSJ1025\_SMG6.1 & 10:25:15.58 & 04:50:25.0 & 9.7 & 3.2374 ± 0.0001 & DG &218 ± 16 & 1.8 ± 0.5 & 5.3 ± 1.5  \\ 
        SDSSJ1025\_SMG11.1 & 10:25:07.55 & 04:52:35.0 & 8.1 & 3.3391 ± 0.0003 & SG &278 ± 50 & 0.7 ± 0.2 & 2.2 ± 0.6 \\ 
        SDSSJ1342\_SMG4.1 & 13:42:18.71 & 16:59:18.5 & 8.0 & 3.1031 ± 0.0002 & SG &399 ± 39 & 1.4 ± 0.5 & 3.8 ± 1.4  \\ 
        SDSSJ1342\_SMG7.1 & 13:42:16.91 & 17:05:37.4 & 8.0 & 3.0724 ± 0.0003 & SG & 261 ± 48 & 0.6 ± 0.2 & 1.6 ± 0.5 \\ 
        \hline
        Supplementary \\
        ID(ALMA, LSB)& RA & Dec & S/N & $z_{spec}$  & model & FWHM & $I_{\rm CO}$ & $L'_{\rm CO(4-3)}\ ^a$ \\
         ~&~&~&~&~&~&[km/s]&[Jykm/s]&[$10^{10} K km s^{-1} pc^{2}$] \\ \hline
        Q2355\_SMG4.1 & 23:58:17.97 & 01:23:43.6 & 7.2 & 4.0302 ± 0.0003 &  DG &222 ± 21 & 1.4 ± 0.5 & 5.8 ± 2.1\\ 
        SDSSJ0819\_SMG5.1 & 08:19:25.59 & 08:22:59.2 & 7.1 & 3.7508 ± 0.0009 & DG &418 ± 79 & 1.0 ± 1.2 & 3.7 ± 4.5 \\ 
        SDSSJ1020\_SMG5.1 & 10:20:08.05 & 10:40:00.4 & 7.6 & 3.7793 ± 0.0004 & DG &225 ± 26 & 0.9 ± 0.4 &  3.4 ± 1.5\\ 
        SDSSJ1025\_SMG10.1 & 10:25:03.74 & 04:54:49.3 & 7.5 & 3.7457 ± 0.0002 & DG &407 ± 12 & 2.7 ± 0.7 & 10.0 ± 2.6 \\ 
        SDSSJ1209\_SMG6.1 & 12:09:24.89 & 11:39:48.7 & 7.2 & 3.6469 ±  0.0002 & DG &225 ± 16 & 1.8 ± 0.5 &  6.4 ± 1.8\\ 
        SDSSJ1209\_SMG13.1 & 12:09:11.11 & 11:40:14.3 & 9.0 & 3.5856 ± 0.0007 & DG &411 ± 37 & 1.0 ± 0.2 & 3.5 ± 0.7\\ 
        \hline\hline
    \end{tabular}}
    \tablecomments{$\ ^a$ Here, we show the CO luminosity of the emission lines which are expected to be CO(4-3) for ALMA. The CO luminosity of other CO transition are mentioned in Appendix \ref{apx:a}.}
    \label{tab:tab3} 
\end{table*}

\subsubsection{Curve-of-Growth Analyses} \label{subsubsec:312}
\begin{figure}[ht!]
\includegraphics[width=8.5cm]{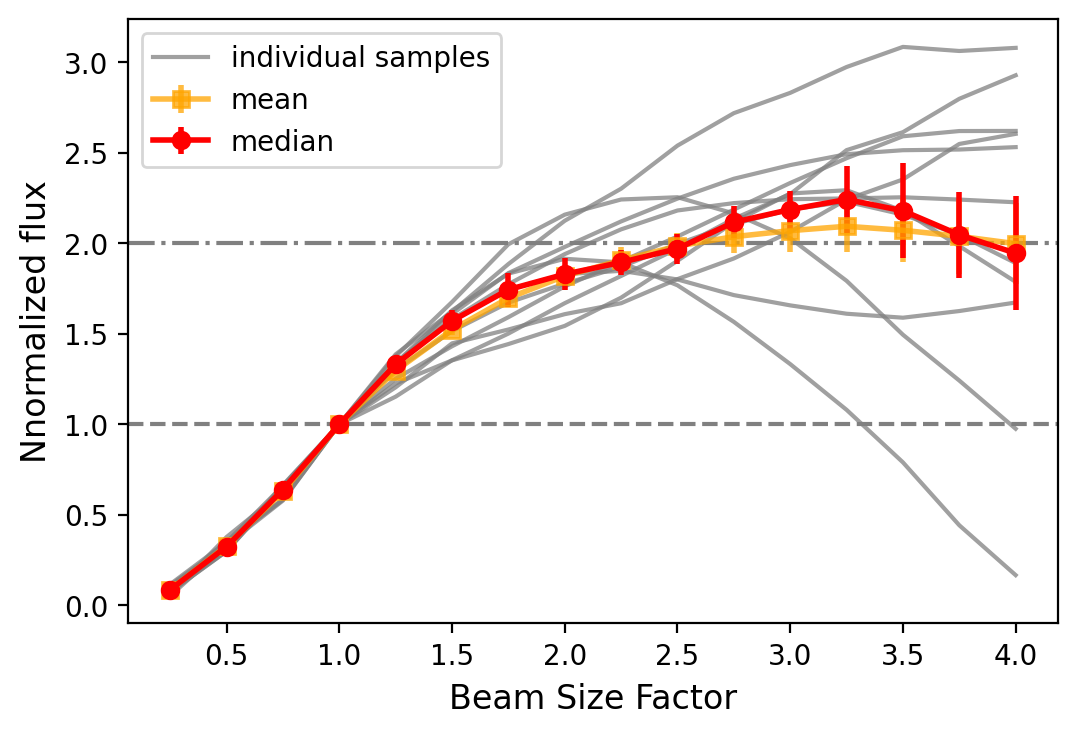}
\centering
\caption{The results of the curve-of-growth analyses. Gray lines represent CO intensity of each sample SMG extracted using various beam sizes, normalized to that of original beam size. The orange and red line represent the mean and median normalized CO intensity value, respectively. Both the mean and median converge to 2.0 at sufficiently large aperture sizes ($\gtrsim$2 beams). 
\label{fig:fig2}}
\end{figure}

For spectral analyses, the extracted flux intensity depends on the size of apertures. To derive the total flux intensity, we follow the curve-of-growth analyses performed in the case with other ALMA data sets \citep{2022ApJ...929..159C}. We varied our aperture from 0.25 to 4.00 times the beam size and repeated all steps of the spectral extraction analyses. The results are shown in \autoref{fig:fig2}, where the analyses are conducted on both the primary and supplementary samples. As anticipated, the curves derived from the samples are expected to converge to a specific level. However, some curves exhibit a downward trend or a monotonically increasing pattern. We suspect that this behavior is attributed to larger-scale noise structures, which have also been observed in previous studies (e.g., \citealp{2020ApJ...904..131N, 2022ApJ...929..159C}).
We then calculated the mean and median values for each beam factor (orange and red curve in \autoref{fig:fig2}). We found that the median/mean line intensity converges to two times the intensity extracted with one beam at beam factors larger than 2.
As a result, we multiplied our intensity results, which were extracted within one beam size, by a factor of 2 to obtain the total flux intensity. 

\subsubsection{Survey Depth Analyses} \label{subsubsec:313}
To determine the SNR threshold for our samples, we perform Monte Carlo simulations to investigate the depths and completeness of the detections. To explain our method, in the following we use the results based on the USB of the ALMA data cubes as an example. First, we generate fake emission lines with a single Gaussian profile with various line widths (FWHM) and line intensity with the same spatial shape as the beam size of the original cube. In this case, we assume these lines to be CO(4-3), as this is the most likely transition for the majority of our detections. Second, we inject five fake lines into random spectral cubes of the same quasar field, at random locations and frequency channels. We repeat this step 10 times, resulting in a total of 50 fake line sources, and perform this process for each quasar field. This step avoids overcrowding of the fake sources in the cube compared to injecting 50 fake sources at once. Then, we perform spectral extraction again on the spectral cubes containing the fake lines, considering different SNR limits. Excluding the real sources in the cubes, we then calculate the completeness (COM) and false detection rate (FDR) defined as 

\begin{displaymath}
\rm COM =\frac{\text{\# of fake lines detected}}{\text{\# of fake lines}}
\end{displaymath}

\begin{displaymath}
\rm FDR =  1- \frac{\text{\# of fake lines detected}}{\text{\# of all lines detected}}
\end{displaymath}

\par
\begin{figure*}[ht!]
\centering
\includegraphics[width=\textwidth]{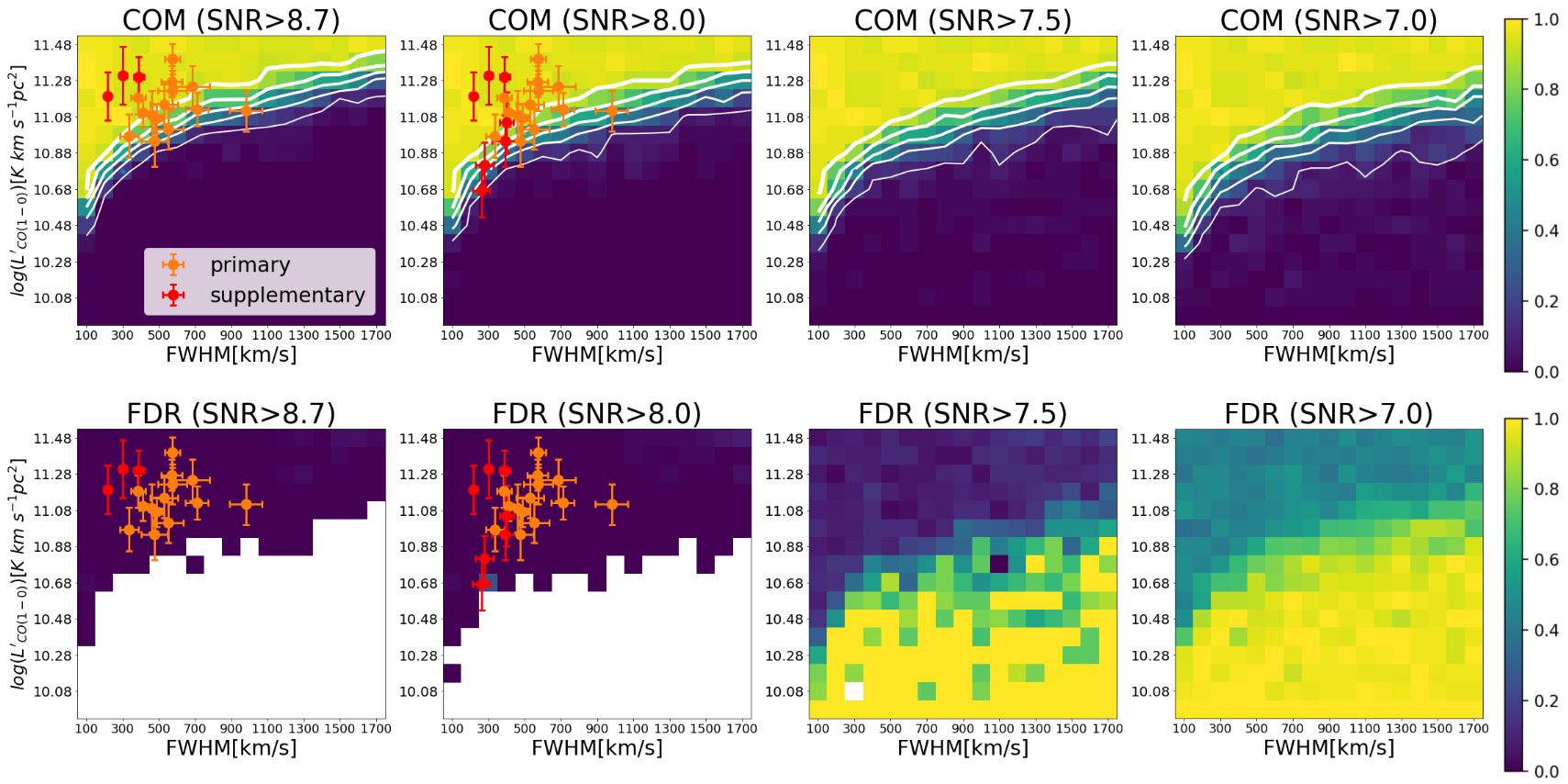}
\caption{Completeness (COM) and false detection rate (FDR) based on our Monte Carlo simulations using ALMA USB cubes. \textbf{Upper panels:} COM with different signal-to-noise ratio (SNR). The corresponding SNRs from the right to left panels are 7.0, 7.5, 8.0, and 8.7 (the minimum SNR among primary sources). The white curves show the completeness at 10\%, 30\%, 50\%, 70\% and 90\% levels from thin to thick. \textbf{Lower panels:} FDR at different SNR cuts. The white pixels represent the locations where no sources can be detected. Additionally, in panels with SNRs $\geq$ 8.0 (our detection threshold), we plot the primary (orange points) and supplementary (red points) sources that exceed the corresponding SNR threshold.
\label{fig:fig3}}
\end{figure*}

The COM and FDR results for USB cubes at different SNR limits are displayed in the upper and lower panels of \autoref{fig:fig3}, respectively. It should be noted that the median value of FDR selected at SNRs over 7.5 is $\sim$ 0.1 for brighter sources, but the FDR for a SNR $\geq$ 8.0 cut is zero for most of the grids. This motivated our choice of the detection threshold (SNR $\geq$ 8.0) for the ALMA USB cubes. 
It is worth noting that the minimum SNR for the primary sources is 8.7, and most of the supplementary sources in the USB have their SNR range from 8.0 to 8.7. 

We now apply the same steps to the LSB of the ALMA data cubes, and 
find that it shows similar results in COM and FDR as those based on the USB data cubes, but with a lower SNR threshold of $\geq$ 7.0. 
Thus, for the selection of LSB sources, we choose the detections whose SNRs is over 7.0. Because of the similar observational conditions 
(e.g., exposure time), the COM and FDR results are similar among each field. Likewise, the same steps are applied to the NOEMA cubes. Due to the similar observational depth, 
we find similar results for COM and FDR as for the ALMA data cubes, and we adopt the same detection thresholds for NOEMA as those for ALMA.

\par 

\subsection{Continuum} \label{subsec:32}
As we describe in Section~\ref{sec:Observation}, we have 
continuum data in various wavelengths for our targets
taken from JCMT/SCUBA-2 at 850\,$\mu$m, 450\,$\mu$m, and ALMA at 3\,mm or NOEMA at 2\,mm. Additionally, we used the public Herschel data for Q2355 and SDSSJ0819 at 250\,$\mu$m, 350\,$\mu$m, 500\,$\mu$m. Using these available data, we obtained the continuum fluxes for our sample sources by searching the peak value within the beam size of the corresponding maps centered at the peak pixels of the CO emission lines. 

However, the low spatial resolution of the Herschel maps makes them prone to blending and overestimating the continuum flux. Due to a lack of high-resolution maps at relevant wavelengths in our target fields, particularly mid-infrared and radio, we estimate the amount of blending using data and catalogs in the well-known COSMOS field. We consider the COSMOS super-deblended catalog reported by \cite{2018ApJ...864...56J}, and Herschel level-2 maps at 250\,$\mu$m, 350\,$\mu$m and 500\,$\mu$m  (HerMES Large Mode Survey; PI: Viero, M.). We apply our method to extract continuum flux at the corresponding locations of the sources in the super-deblended catalog, and calculate the median ratio between the extracted flux and the deblended flux provided in the catalog. The same steps are applied to the SCUBA-2 maps to obtain the deblending ratios at 450\,$\mu$m and 850\,$\mu$m \citep{2024ApJ...971..117G}. As a result, we obtain the deblending ratios ranging from $\sim$0.7 - 1.0, which are then applied to the extracted fluxes of our sample. The de-blended continuum fluxes at each wavelength are shown in \autoref{tab:tab4}. These continuum data are used to verify the robustness of the line emitters as well as to enable us to conduct SED fittings. 
\par

\begin{table*}[h]
    \centering
    \caption{Continuum of CO emitters at far-infrared 
    wavelengths}
    \setlength{\tabcolsep}{2mm}{
    \begin{tabular}{lllllll}
    \toprule
        Primary \\
        ID & ALMA  & SCUBA-2$^*$  & ~  & Herschel$^*$ & ~ & ~\\
        ~ & 3\,mm [mJy]  & 850\,$\mu$m [mJy]  & 450\,$\mu$m [mJy]  & 250\,$\mu$m [mJy] & 350\,$\mu$m [mJy] & 500\,$\mu$m [mJy] \\ \hline
        
        Q2355\_SMG8.1 & 0.07 ± 0.04 & 7.3 ± 1.6 & 89.6 ± 28.4 & 3.6 ± 21.8 & 34.3 ± 10.3 & 6.2 ± 20.3 \\
        Q2355\_SMG9.1 & 0.23 ± 0.04 & 6.4 ± 1.5 & 27.3 ± 23.3 & - & 7.2 ± 12.2 & 33.0 ± 13.7 \\
        SDSSJ0819\_SMG1.1 & 0.33 ± 0.05 & 14.0 ± 1.2 & 45.6 ± 12.2 & 25.3 ± 10.1 & 33.4 ± 5.4 & 41.8 ± 6.3 \\
        SDSSJ0819\_SMG2.1 & 0.27 ± 0.05 & 14.5 ± 1.8 & 58.9 ± 19.6 & 58.1 ± 9.0 & 59.1 ± 6.7 & 65.2 ± 7.9 \\
        SDSSJ0819\_SMG3.1 & 0.19 ± 0.05 & 12.2 ± 1.6 & 43.0 ± 17.5 & 30.3 ± 15.1 & 27.6 ± 6.5 & 19.4 ± 11.3 \\
        SDSSJ0819\_SMG4.1 & 0.27 ± 0.05 & 9.6 ± 1.6 & 35.9 ± 20.6 & 25.5 ± 11.6 & 14.0 ± 9.6 & 30.3 ± 16.4 \\
        SDSSJ0819\_SMG10.1 & 0.05 ± 0.05 & 6.4 ± 1.4 & 30.8 ± 15.8 & 19.9 ± 8.8 & 5.6 ± 9.5 & 5.2 ± 9.9 \\
        SDSSJ0819\_SMG15.1 & 0.18 ± 0.05 & 8.6 ± 2.0 & 51.6 ± 23.1 & 17.7 ± 14.6 & 0.4 ± 15.8 & - \\
        SDSSJ1020\_SMG3.1 & 0.07 ± 0.05 & 8.8 ± 1.0 & 19.7 ± 15.5 & - & - & - \\
        SDSSJ1025\_SMG1.1 & 0.09 ± 0.04 & 9.8 ± 1.2 & 29.8 ± 10.6 & - & - & - \\
        SDSSJ1209\_SMG2.1 & 0.21 ± 0.05 & 7.6 ± 0.7 & 17.1 ± 8.5 & - & - & - \\
        SDSSJ1209\_SMG4.1 & 0.16 ± 0.05 & 7.5 ± 1.1 & 44.0 ± 13.0 & - & - & - \\
        SDSSJ1342\_SMG1.1 & 0.18 ± 0.03 & 9.6 ± 1.0 & 24.6 ± 12.3 & - & - & - \\
        SDSSJ1342\_SMG5.1 & 0.15 ± 0.03 & 7.0 ± 1.3 & 43.9 ± 16.4 & - & - & - \\ \hline

        Primary \\
        ID & NOEMA  & SCUBA-2$^*$  & ~  & Herschel$^*$ & ~ & ~\\
        ~ & 3\,mm [mJy]  & 850\,$\mu$m [mJy]  & 450\,$\mu$m [mJy]  & 250\,$\mu$m [mJy] & 350\,$\mu$m [mJy] & 500\,$\mu$m [mJy] \\ \hline
        Jackpot\_SMG5.1 & 0.12 ± 0.01 & 7.3 ± 0.8 & 17.2 ± 5.2 & - & - & - \\ 

        \hline\hline
        Supplementary \\
        
        ID & ALMA(USB)  & SCUBA-2$^*$  & ~  & Herschel$^*$ & ~ & ~\\
        ~ & 3\,mm [mJy]  & 850\,$\mu$m [mJy]  & 450\,$\mu$m [mJy]  & 250\,$\mu$m [mJy] & 350\,$\mu$m [mJy] & 500\,$\mu$m [mJy] \\ \hline
        SDSSJ1020\_SMG10.1 & -0.03 ± 0.10 & 0.7 ± 0.5 & 24.0 ± 10.6 & - & - & - \\ 
        SDSSJ1025\_SMG4.1 & 0.06 ± 0.13 & 0.3 ± 1.1 & 15.0 ± 12.1 & - & - & - \\ 
        SDSSJ1025\_SMG4.2 & 0.05 ± 0.05 & 3.8 ± 1.3 & 26.6 ± 13.8 & - & - & - \\ 
        SDSSJ1025\_SMG6.1 & 0.08 ± 0.10 & 2.8 ± 1.2 & 33.6 ± 12.4 & - & - & - \\ 
        SDSSJ1025\_SMG11.1 & 0.02 ± 0.04 & 2.2 ± 0.9 & 18.9 ± 7.9 & - & - & - \\ 
        SDSSJ1342\_SMG4.1 & -0.03 ± 0.07 & 3.6 ± 1.4  & 30.8 ± 21.0  & - & - & - \\ 
        SDSSJ1342\_SMG7.1 & 0.01 ± 0.03 & 5.0 ± 1.2 &  51.5 ± 17.5 & - & - & - \\ 
        
        \hline
        Supplementary \\
        ID & ALMA(LSB)  & SCUBA-2$^*$  & ~  & Herschel$^*$ & ~ & ~\\
        ~ & 3\,mm [mJy]  & 850\,$\mu$m [mJy]  & 450\,$\mu$m [mJy]  & 250\,$\mu$m [mJy] & 350\,$\mu$m [mJy] & 500\,$\mu$m [mJy] \\ \hline
        Q2355\_SMG4.1 & 0.32 ± 0.10 & 3.1 ± 1.5 & 42.3 ± 21.1 & -4.3 ± 24.5 & -14.4 ± 24.6 & -14.0 ± 11.7 \\ 
        SDSSJ0819\_SMG5.1 & 0.11 ± 0.05 & 3.6 ± 1.4 & 44.2 ± 17.8 & 14.0 ± 8.0 & 4.1 ± 22.5 & 7.1 ± 11.2 \\ 
        SDSSJ1020\_SMG5.1 & 0.82 ± 0.11 & 1.6 ± 0.5 & 17.5 ± 9.8 & - & - & - \\ 
        SDSSJ1025\_SMG10.1 & 0.20 ± 0.16 & 1.9 ± 1.2 & 9.8 ± 10.6 & - & - & - \\ 
        SDSSJ1209\_SMG6.1 & 0.06 ± 0.08 & 0.3 ± 0.9 & 24.1 ± 10.6 & - & - & - \\ 
        SDSSJ1209\_SMG13.1 & 0.06 ± 0.06 & 4.8 ± 0.9 & 27.6 ± 11.2 & - & - & - \\ \hline\hline

    \end{tabular}}
    \tablecomments{ $\ ^*$Deblended continuum flux.}
    \label{tab:tab4} 
\end{table*}

\subsubsection{SED Fitting}\label{sec:sed}
To model the SED of the detected 
sources, 
we adopt a single temperature modified black-body (MBB) as usually done in the literature \citep{2012MNRAS.425.3094C, 2021ApJ...919...30D}, defined as:

\begin{equation}
S_\nu \propto [1-exp(-\tau_\nu)]B_\nu(T_{dust})
\label{eq:eq1} 
\end{equation}
where $B_\nu(T_{dust})$ is the Planck function, and $\tau_\nu$ is the optical depth, defined as: 

\begin{displaymath}
\tau_\nu = \kappa_\nu\Sigma_{dust}
\end{displaymath}
where $\kappa_\nu\Sigma_{dust}$ is the dust mass surface density, and $\kappa_\nu$ is the frequency-dependent dust opacity, defined as: 

\begin{displaymath}
\kappa_\nu \propto (\frac{\nu}{\nu_0})^\beta
\end{displaymath}
where $\beta$ is the dust emissivity index and $\nu_0$ is the reference frequency.
\par
In our case, for simplicity, we assume an optically thin case, with $\tau_\nu \ll 1$. Therefore,  $[1-exp(-\tau_\nu)]\cong\tau_\nu$, so Equation \ref{eq:eq1} becomes

\begin{displaymath}
S_\nu \propto \tau_\nu B_\nu(T_{dust}) = A M_{dust} \nu^\beta B_\nu(T_{dust})
\end{displaymath}
Here, we assume that coefficient A combines all constants. Integrating the model from 8\,$\mu m$ to 1000\,$\mu m$ in wavelength, we can derive the infrared luminosity, $L_{\rm IR}$, which is shown in \autoref{tab:tab6}. We simply fixed $\beta = 2.0$, 
representing the usual condition for dusty galaxies (e.g.,
\citealt{2021ApJ...923..215C,2021ApJ...919...30D,Cooper:2022aa,2024ApJ...961..226L}). An example best-fit result is shown in \autoref{fig:sedfit} and that for all the primary sources are given in Appendix~\ref{apx:f}. The relevant parameters from the best-fit models are also provided in Table~\ref{tab:tab6}. The median values of $T_{\rm dust}$ and ${\rm log}(L_{\rm IR}/[L_\odot])$ are $34 \pm 3\ K$ and $12.8 \pm 0.1$, respectively.


\begin{figure}[ht!]
\centering
\includegraphics[width=8.5cm]{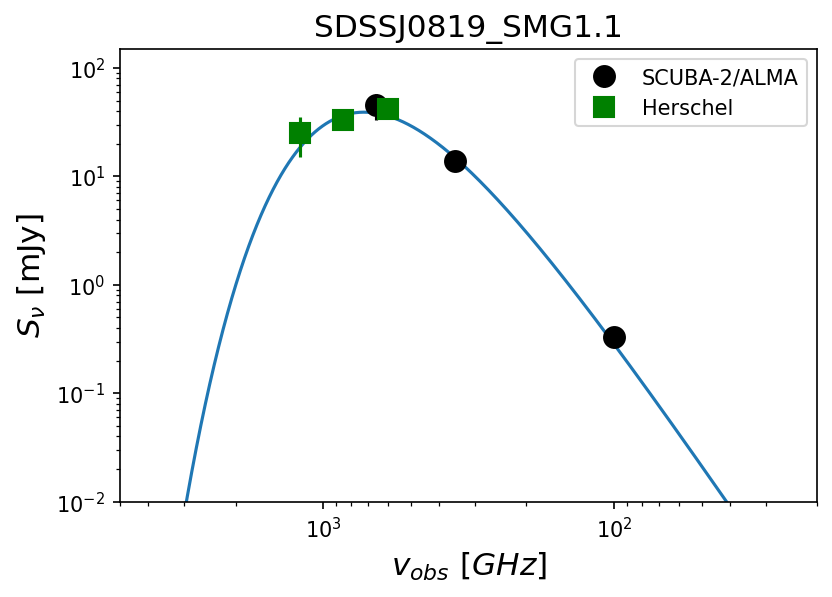}
\caption{SED for SDSSJ0819\_SMG1.1: One example of the MBB fitting results for the primary sample. Black points represent the flux density for ALMA and JCMT/SCUBA-2 (from left to right: 450\,$\mu m$, 850\,$\mu m$ (SCUBA-2), 3\,mm (ALMA)), while green points represent those for Herschel (from left to right: 250\,$\mu m$, 350\,$\mu m$, 500\,$\mu m$). The blue line represents the fitting result. The fitting results of the rest of the sample can be found in Appendix~\ref{apx:f}.
\label{fig:sedfit}}
\end{figure}

\begin{table*}[t]
    \centering
    \caption{MBB fitting results for primary samples}
    \setlength{\tabcolsep}{3mm}{
    \begin{tabular}{lcccc}
    \toprule
        ID & $T_{\rm dust}\ ^a\ $ &$ M_{\rm dust}\ ^b\ $ & $log(L_{\rm IR})^c$ & $\beta \,^d$ \\ \hline
        Q2355\_SMG8.1 & 39 ± 5 & 1.0 ± 0.4 & 13.0 ± 0.3 & 2.7 ± 0.7\\
        Q2355\_SMG9.1 & 24 ± 3  & 4.5 ± 1.5 & 12.3 ± 0.4  & 1.4 ± 0.5\\
        SDSSJ0819\_SMG1.1 & 29 ± 1 & 4.6 ± 0.7 & 12.9 ± 0.1 & 1.8 ± 0.2\\
        SDSSJ0819\_SMG2.1 & 35 ± 1 & 3.1 ± 0.5 & 13.2 ± 0.1 & 1.9 ± 0.2\\
        SDSSJ0819\_SMG3.1 & 30 ± 2 & 3.3 ± 0.7 & 12.8 ± 0.2 &  2.1 ± 0.3\\
        SDSSJ0819\_SMG4.1 & 27 ± 2 & 4.3 ± 1.2 & 12.6 ± 0.3 &  1.5 ± 0.3\\
        SDSSJ0819\_SMG10.1 & 33 ± 4 & 1.2 ± 0.5 & 12.6 ± 0.4 & 2.4 ± 0.8\\
        SDSSJ0819\_SMG15.1 & 28 ± 4 & 3.1 ± 1.2 & 12.6 ± 0.4 & 2.0 ± 0.5\\ 
        SDSSJ1020\_SMG3.1 & 38 ± 10 & 1.4 ± 0.9 & 13.0 ± 0.8 & -\\ 
        SDSSJ1025\_SMG1.1 & 39 ± 7 & 1.4 ± 0.6 & 13.1 ± 0.5 &-\\ 
        SDSSJ1209\_SMG2.1 & 24 ± 3 & 4.4 ± 1.9 & 12.3 ± 0.4 &-\\ 
        SDSSJ1209\_SMG4.1 & 42 ± 11 & 1.1 ± 0.6 & 13.2 ± 0.7 & -\\ 
        SDSSJ1342\_SMG1.1 & 29 ± 4 & 3.1 ± 1.0 & 12.7 ± 0.4 &-\\ 
        SDSSJ1342\_SMG5.1 & 32 ± 6 & 2.0 ± 0.9 & 12.8 ± 0.6 &-\\
        Jackpot\_SMG5.1 & 30 ± 2 & 2.0 ± 0.3 & 12.6 ± 0.2 &- \\
        \hline\hline
    \end{tabular}}
    \tablecomments{$\ ^a$In unit K.$\ ^b$In unit $10^9 M_{\odot}$.$\ ^c$$L_{\rm IR}$ is in unit $L_{\odot}$. $\ ^d$ Results for free-$\beta$ fitting. \\ Note that the first 3 columns are derived by assuming $\beta$ fixed by 2.0.}
    \label{tab:tab6} 
\end{table*}

\subsection{Detection rate}
\label{subsec:33}


Of the 101 SCUBA-2 sources targeted by our ALMA and NOEMA observations, we detect 28 line-emitting sources in total, and 15 are confirmed physically associated with their corresponding quasars. We stacked the Herschel maps, at the three wavelengths used, on the SCUBA-2 sources that do not have line detection, which yielded continuum detection with significant SNRs (4-6$\sigma$ among three wavelengths). This suggests that the original SCUBA-2 detections are robust as expected. Therefore, the relatively low detection rate merits further discussion.

We start by estimating the expected number of detection in the following. First, based on the 
 number count measurements reported by \citet{FAB2023}, we calculate the excess number of sources relative to the field \citep{2017MNRAS.465.1789G} based on the survey area. We limit the calculation to the flux regime greater than 5 mJy at 850$\,\mu m$, which is our original selection for ALMA and NOEMA follow-up observations.
Next, we consider the following two observational biases. 

1. Spectral coverage ratio (SCR), which is caused by insufficient spectral coverage. 
As we mentioned in Section \ref{subsubsec:311}, primary samples lie within the velocity distribution range within 7000 km/s to ensure that they are members relevant to the large-scale structure of the protocluster.
However, because of the tuning of the ALMA observations, the spectral coverage (e.g., -2000--8000 km/s for USB from ALMA) cannot completely cover the whole distribution. Therefore, the SCR represents the fraction of the reported velocity probability distribution that is covered by our ALMA and NOEMA spectral windows.


2. Detection bias ratio (DBR), which is caused by limited sensitivity. As detailed in Section \ref{subsubsec:313}, we were unable to detect any sources below the sensitivity limit (shadow region in left panel of Figure~\ref{fig:fig7}). Based on the distribution of luminosity and line width of SMGs in the field \citep{2021MNRAS.501.3926B}, we calculate the fraction of sources that would have been detected given our sensitivity limits. As a result, the DBRs are $\sim$81\% and $\sim$77\% for CO(4-3) and CO(3-2), respectively.


\begin{figure}[h]
\centering
\includegraphics[width=8.5cm]{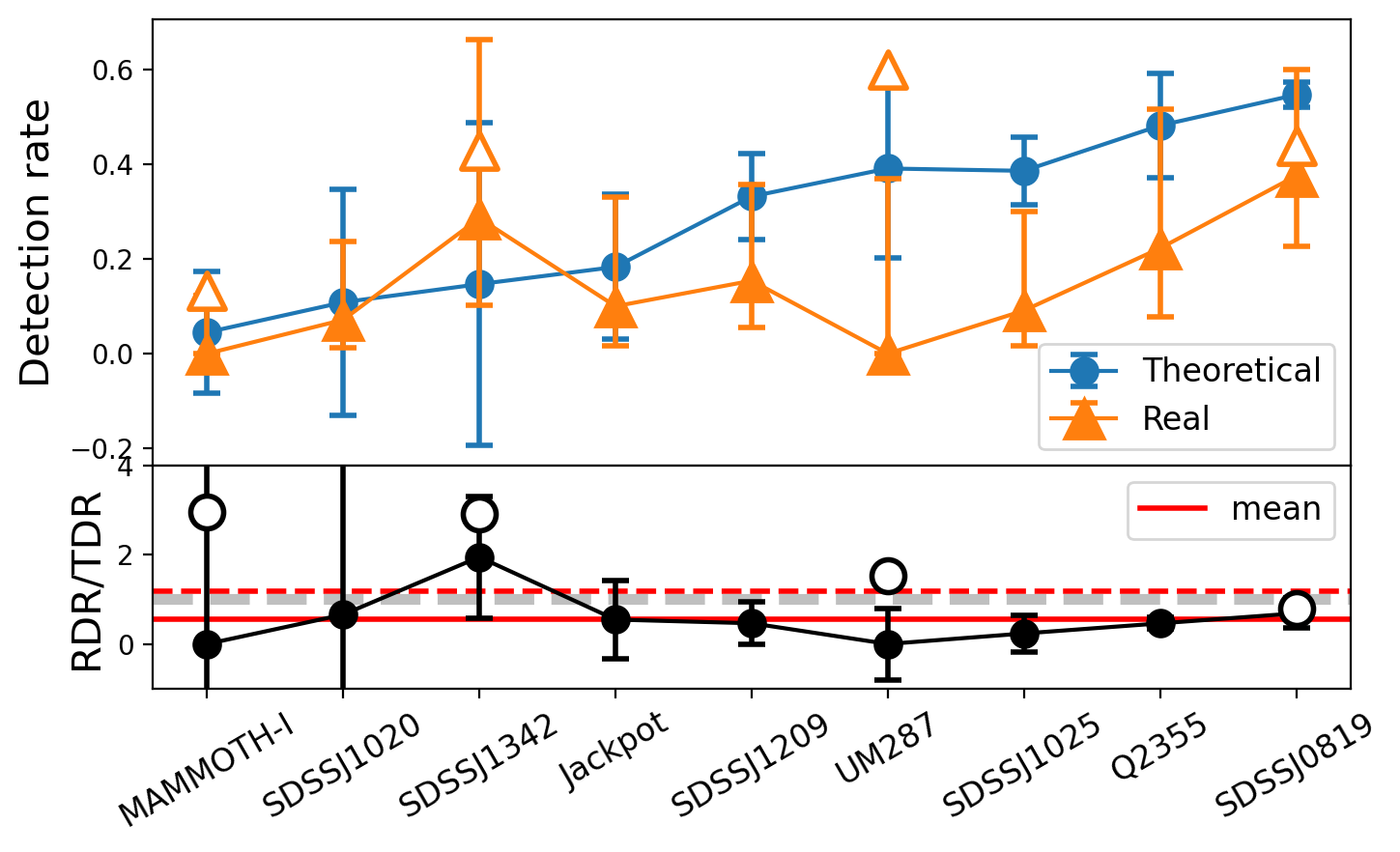}
\caption{ \textbf{Upper panel:} Expected detection rates (blue circles) and observed detection rates (orange triangles) for each quasar field, while the hollow one show the observed rates included low-SNR sources. \textbf{Lower panel:} Ratios between the observed and expected detection rates, where the red solid line represents the mean value, which is about 0.5. The hollow points represent ratio considering low-SNR sources, with the mean value is shown as red dotted line ($\sim$ 1.1).
\label{fig:dtr}}
\end{figure}

\begin{figure*}[ht!]
\centering
\includegraphics[width=\textwidth]{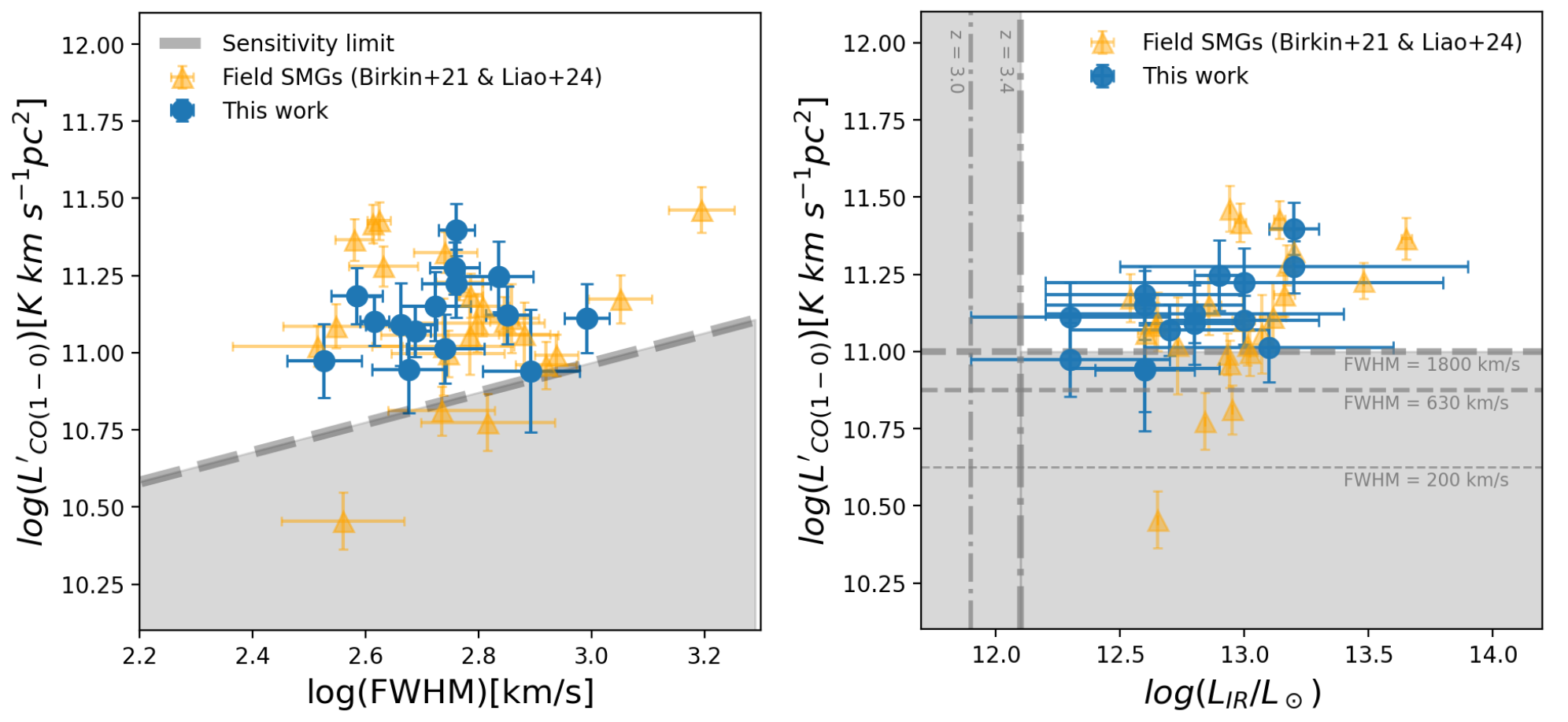}
\caption{\textbf{Left panel:}  $L'_{\rm CO}$-FWHM relation. The blue points represent our primary sample at $z\sim2-3$ from SMGs around quasars. Orange triangles show typical field SMGs from the survey program by \citet{2021MNRAS.501.3926B} and \citet{2024ApJ...961..226L}. Additionally, the gray shadow region shows the detection limit ($\geq$ 8.7 SNRs, the minimum value among primary sources) in our observations derived from Section \ref{subsubsec:313}. 
\textbf{Right panel:} $L'_{\rm CO}$-$L_{\rm IR}$ correlation. The same color and symbols represent the same samples as in the left panel. The gray shadow region shows the detection limits of $L_{\rm IR}$ (vertical dashed lines, corresponding to different redshift assumptions) and $L'_{\rm CO}$ (horizontal dashed lines, corresponding to different FWHM values).
\label{fig:fig7}}
\end{figure*}

We then derive the expected detection rates by multiplying the excess number of sources deduced from the number counts by SCR and DBR, and dividing the results by the total number observed. We compare the expected detection rates with the observed detection rates, which are calculated based on the number of primary sources in the catalog (Table~\ref{tab:tab3}) divided by the number of targets in each quasar field. The results are shown in Figure~\ref{fig:dtr}, where we find that the observed detection rates follow the trend of the expected detection rates. However, the observed detection rates are systematically lower by a factor of two on average compared to the expected rates. Since the adopted velocity range is in principle sufficiently wide to capture galaxies that are located within the same large-scale structures ($\sim \pm$90\,cMpc), the systematically lower detection rates most likely suggest that there is a significant fraction of sources that are associated with the corresponding quasars but lie below our current detection limits. Since there could also be a detection bias in the field SMGs, we refrain ourselves from concluding that this is evidence that SMGs in dense environments have lower CO line luminosity than those in the field. In order to make progress, more studies with deeper spectral line observations are needed for both SMGs in the field and those in dense environments.

To estimate how many sources we missed because of the current detection limit, we search for fainter sources under our SNR threshold (e.g, 8 for USB). Specifically, we search for sources whose SNR is at least 
5.0, but satisfy the conditions: (1) source location within the beam of the 850$\,\mu m$ SCUBA-2 image; (2) FWHM of lines $\geq$ the minimum ($\sim$180 km/s) of the reported lines for literature SMGs \citep{2021MNRAS.501.3926B}; (3) fidelity $\geq$ 0.5, with fidelity defined as 
\begin{displaymath}
\rm fidelity = 1- \frac{N_{neg}(S/N)}{N_{pos}(S/N)}
\end{displaymath}
which depends on the number of negative and positive detections at corresponding SNR. Following these criteria, we found 7 additional detections. The detection rates including these lower-SNR sources are shown as hollow symbols in Figure~\ref{fig:dtr}. When considering the additional sources, the average ratio between real and expected rate approaches unity. 
The comparable level between real and expected rate suggests that some fainter sources are hidden in the noise. 

In summary, we find that our detection rates can be explained by jointly considering the original SCUBA-2 number counts, the incompleteness of the spectral coverage, and that about half of the physically associated sources lie below the current detection limit.


\section{Discussion} \label{sec:Discussion}

\subsection{Molecular Gas and Dust Properties} \label{subsec:41}
With the deduced gas and dust properties of our sample in the dense environment, we are now ready to compare our results to similar sources in the field, in order to understand if environment impacts significantly on these properties. In the following sections we made comparisons with mainly three field SMG samples \citep{2020MNRAS.494.3828D,2021MNRAS.501.3926B,2024ApJ...961..226L}, which are all ALMA follow-up of 850/870\,$\mu$m selected sources from single-dish SCUBA-2 or LABOCA surveys with similar depths of our survey, with their $\rm S_{870\mu m}$ ranging from 2 to 20 mJy. Note that unlike the other field SMG samples analyzed in this study, the sources presented in \citet{2024ApJ...961..226L} are significantly brighter at submillimeter wavelength ($\rm S_{870\mu m} = 12.4 - 19.2\,\rm mJy$), and can therefore be categorized as bright field SMGs. We also caution that although the samples reported by \citet{2021MNRAS.501.3926B} are often considered representative of typical field SMGs, they exhibit a systematically higher $\rm S_{870\mu m}$ than broader samples reported by \citet{2020MNRAS.494.3828D}, leading to a systematic offset of approximately 0.2 dex in their ISM properties.

\subsubsection{CO line properties}

 We first examine the relationship between CO luminosity and the line width (FWHM) of the emission lines in the left panel of Figure~\ref{fig:fig7}. Since CO luminosity is considered a tracer of the molecular gas reservoir \citep{Solomon:2005aa} and FWHM reflects the gas dynamics, this relation probes the correlation between gas mass and dynamics \citep{2012ApJ...752..152H}. The median FWHM value for our samples is 498 $\pm$ 36 km\,s$^{-1}$, comparable to the median value 569 $\pm$ 47 km\,s$^{-1}$ reported for the field SMGs \citep{2021MNRAS.501.3926B}. After accounting for the sensitivity limits, we conclude that there is no significant difference in CO line profiles between our samples and field SMGs.

 Next, we look at the relationship between CO luminosity and infrared luminosity. In general, 
infrared luminosity is considered a tracer of ongoing star formation, averaging over the past $\sim10-100$\,Myr \citep{1998ARA&A..36..189K}. Therefore, the relation between $L'_{\rm CO}$ and $L_{\rm IR}$ can be interpreted as a measure of the fraction of total molecular gas in galaxies that is being actively used for star formation, commonly referred to as the star formation efficiency.

We present the results of the $L'_{\rm CO}$-$L_{\rm IR}$ relation for our primary sources in the right panel of Figure~\ref{fig:fig7}. Concurrently, we also display results from the literature based on the latest CO surveys of the same CO transition for SMGs in the field \citep{2021MNRAS.501.3926B, 2024ApJ...961..226L}. In this work, the median value of $L'_{\rm CO(1-0)}$ is (11.7 ± 1.5) $\times 10^{10}$\,K\,km\,s$^{-1}$\,pc$^2$, whereas the median of log$(L_{\rm IR}/L_{\odot})$ is 12.8 ± 0.1 dex. Comparatively, 
the median values for the field SMGs are $L'_{\rm CO(1-0)}=(11.0\pm0.1)\times 10^{10}$\,K\,km\,s$^{-1}$\,pc$^2$ and log$(L_{\rm IR}/L_{\odot})=12.9\, \pm \, 0.1$\,dex, respectively.
It is evident that our sample sources follow a distribution similar to that of the field SMGs within uncertainties. We also estimate and present the $L_{\rm IR}$ limits by adopting the average 3$\sigma$ upper limits from the 850\,$\mu$m, 450\,$\mu$m, and 3\,mm maps, and performing MBB fitting under the assumption of different redshifts across our sample. However, uncertainties in the assumed redshifts and SED shapes introduce an uncertainty of approximately $\pm$0.2 dex in $\log(L_{\rm IR})$. Therefore, these values should be considered indicative only.

In summary, the CO line properties of our sample SMGs are consistent with those in the field, in agreement with the recent findings by \citet{2017A&A...608A..48D}, who compared CO properties between protocluster and field samples.

\subsubsection{Gas mass fraction} \label{subsec:412}
\begin{figure}[h!]
\centering
\includegraphics[width=8.5cm]{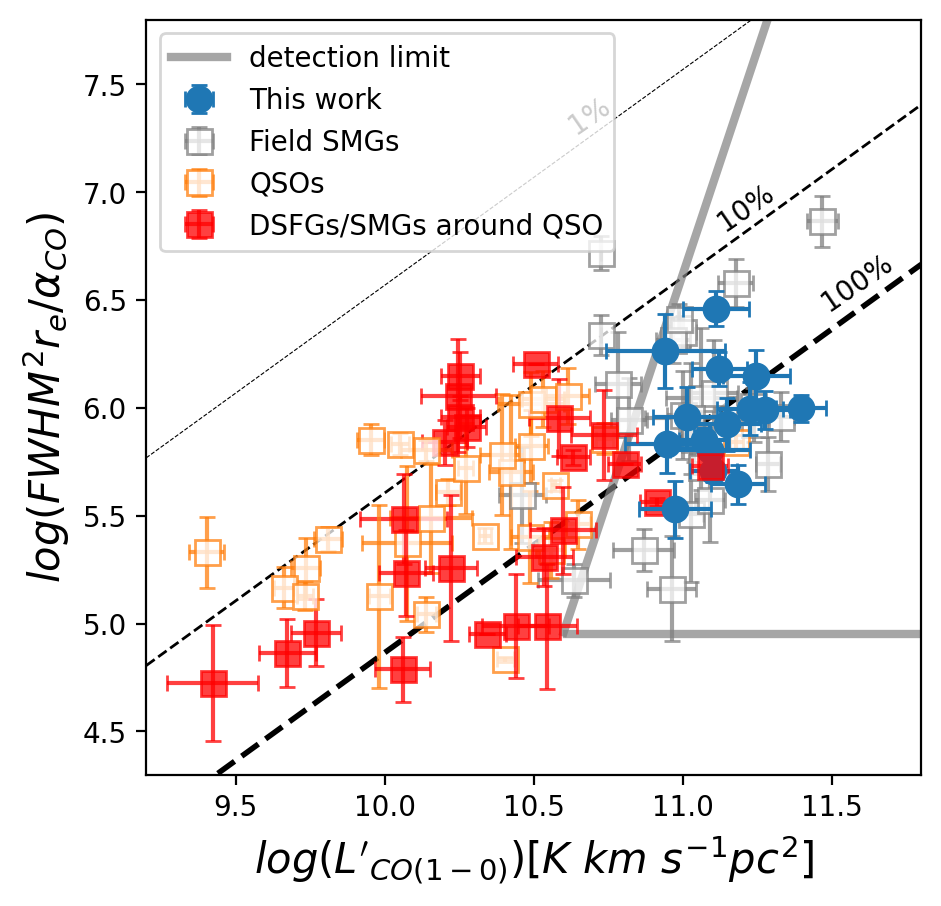}
\caption{A diagram that visualizes gas fractions using a dynamical mass method. The diagonal lines represent the gas mass fraction $f_{gas}$ = 1\%, 10\%, and 100\%, from left to right. The QSO samples \citep{2013ARA&A..51..105C,2021A&A...645A..33B, 2021ApJ...923..200C,  Li2023} and field SMG samples \citep{2013MNRAS.435.1493A, 2013MNRAS.429.3047B, 2021MNRAS.501.3926B} are shown as orange points and gray points, respectively. These two types of samples separate into two distributions; QSOs have lower $f_{gas}$, and field SMGs have higher $f_{gas}$. Blue points represent our SMGs sample, which is at a Mpc scale distance from the central QSOs. 
Red points denote DSFGs/SMGs around $z\sim2-3$ QSO at a distance ranged from a few hundred kpc to Mpc scale \citep{2021ApJ...923..200C,2022ApJ...930...72A, Garcia2022, Li2023, 2024A&A...684A.119P, 2024arXiv240616637W}. Gray lines outline the detection limits of our ALMA observations.
\label{fig:fig8}}
\end{figure}

The gas mass fraction ($f_{gas}$), defined as the ratio between gas mass and stellar mass, is one key measure to constrain the relative amount of fuel for galaxies to feed their star formation \citep{2020ARA&A..58..157T}. The lack of optical and infrared data means that direct estimates of stellar masses for our sources are not yet possible. To bypass this difficulty, we adopt a method that utilizes the dynamical mass and gas mass estimates enabled by our data. In particular, the following formula proposed by \citet{2021ApJ...923..200C} is used: 
\begin{equation}\label{eqn:fgas}
\begin{aligned}
&{\rm log}(\frac{FWHM_{\rm CO}^2r_{e}}{\alpha_{\rm CO}}) = \\
&{\rm log}(L'_{\rm CO(1-0)})-{\rm log}(\frac{3.1\times10^5f_{gas}(1-f_{\rm DM})}{1+f_{gas}})
\end{aligned}
\end{equation}
The formula assumes dynamical equilibrium of rotation-dominated disks, which is consistent with recent kinematic studies that have found high V/$\sigma$ values for dusty star-forming galaxies at cosmic noon (e.g., \citealt{Rizzo:2023aa, Amvrosiadis:2025aa}). Following \citet{2021ApJ...923..200C}, we set the half-light radius to $r_{e} $= 3 kpc, $\alpha_{\rm CO} $= 1.0, and dark matter fraction as $f_{\rm DM}$ = 0.12, based on recent measurements.
With this formula, we estimate $f_{gas}$ from the measured line width FWHM and the CO line luminosity. While the dynamical method has been tested and shown to agree with the direct method when stellar mass estimates are available, the unknown inclination angle of the disks inevitably introduces large scatter of the estimates. We therefore focus primarily on the average trend rather than on individual source values.

We present the results based on the measurements of our primary sources as well as those from the literature in Figure~\ref{fig:fig8}, illustrating the correlation between log$(\rm FWHM_{CO}^2r_{e}/\alpha_{\rm CO})$ and log$(L'_{\rm CO(1-0)})$, with the various values of $f_{gas}$ plotted as diagonal lines.  
Interestingly, our sample SMGs, located at Mpc-scale distances 
from the central quasars, have similar $f_{gas}$ as the field SMGs. 
Additionally, we included some ALMA-identified DSFG/SMGs samples located in the vicinity of quasars with distance ranged from hundred kpc to Mpc scales \citep{2021ApJ...923..200C, 2022ApJ...930...72A, Garcia2022, Li2023, 2024A&A...684A.119P, 2024arXiv240616637W}\footnote{Lines that are better fit by double Gaussian would have their line widths overestimated using single Gaussian models. As most works only reported fitting results using single Gaussian model, we only include lines that appear single peak or could be better fit by single Gaussian models.}.
Intriguingly, some of DSFGs/SMGs appear to have lower $f_{gas}$, aligning more closely with their host quasars. These results may suggest that the distance from the central quasar influences the gas fraction of DSFGs/SMGs.

\begin{figure}[h]
\centering
\includegraphics[width=8.5cm]{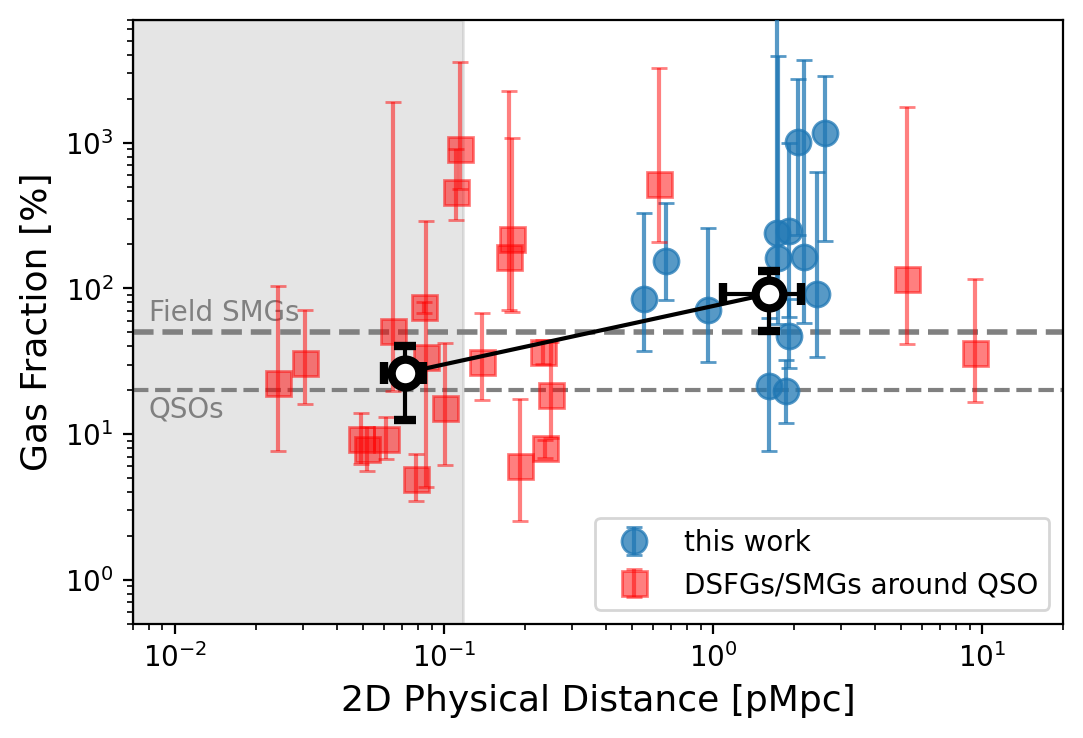}
\caption{Gas fraction as a function of projected distance from the central quasars. Red points represent literature DSFGs/SMGs that are located around QSOs at $z\sim2-3$ \citep{2021ApJ...923..200C,2022ApJ...930...72A, Garcia2022, Li2023, 2024A&A...684A.119P, 2024arXiv240616637W}. Blue points depict our sample SMGs. The black points denote the binned median values (inside and outside the $R_{vir}$), with uncertainties calculated from bootstrapping. The shadowed region represents the range within $R_{vir}$ for $ M_{\rm halo} = 10^{12.5} M_{\odot}$ at $z\sim$3. The upper dashed line represents the median value of field SMGs samples, and lower one represents the median value of QSOs samples.
\label{fig:fig9}}
\end{figure}

To further investigate this suggestion, we calculated the gas mass fraction for both our sample SMGs and literature samples around quasars using Equation \ref{eqn:fgas} with the aforementioned assumptions, then depicted the distribution as a function of their projected distances from the central quasars in Figure~\ref{fig:fig9}. It becomes evident that $f_{gas}$ appears lower within the virial radius $R_{vir}$ ($\sim$100-200\,kpc at $z\sim2-3$) of the quasar halo assuming $M_{\rm halo}\sim10^{12.5} M_{\odot}$, with a mdeian $f_{gas}$ of 26 $\pm$ 13 (comparable to the median value for quasars of 20 $\pm$ 10). The assumed quasar halo mass is supported by clustering measurements (e.g., \citealp{2017MNRAS.468..728R}
) and by Ly$\alpha$ emitters overdensities and kinematics on scales of hundreds of kpc (\citealp{2021MNRAS.503.3044F}). Beyond $R_{vir}$, the $f_{gas}$ of DSFGs (median of 92 $\pm$ 40) is comparable to that of field SMGs (median of 50 $\pm$ 25).

In Figure~\ref{fig:fig8} we plot the detection limits of our data.
Considering the largest observed line width (FWHM of $\sim$1500 km/s) among CO emitters from field SMGs \citep{2021MNRAS.501.3926B}, we can in principle probe gas fractions down to 10\% at a fixed ${\rm log}(L'_{\rm CO(1-0)})$, $\sim$11.0-11.5 in this case. However, most sources exhibit gas fractions close to 100\%, suggesting that our sample is characterized by intrinsically high gas fractions. 
 We caution that the limited sensitivity of our ALMA data prevents us from detecting sources with CO luminosity comparable to those around quasars. Whether these fainter CO sources that are located outside the virial radius have gas fractions as high as those with higher CO luminosity remains to be tested with deeper observations in the future. On the other hand, sources in the literature studies are detected by much deeper data. We therefore expect the detection limits of previous studies to be sloped similarly to the diagonal limit line in our figure, but shifted toward the lower $L^\prime_{\rm CO}$ in the x-axis. In addition, no FWHM threshold was imposed on source detection in previous studies, which means that there would not be a horizontal cutoff on the y-axis. Therefore, we do not expect previous studies to miss sources with high gas fractions.



\begin{figure}[ht!]
\centering
\includegraphics[width=8.5cm]{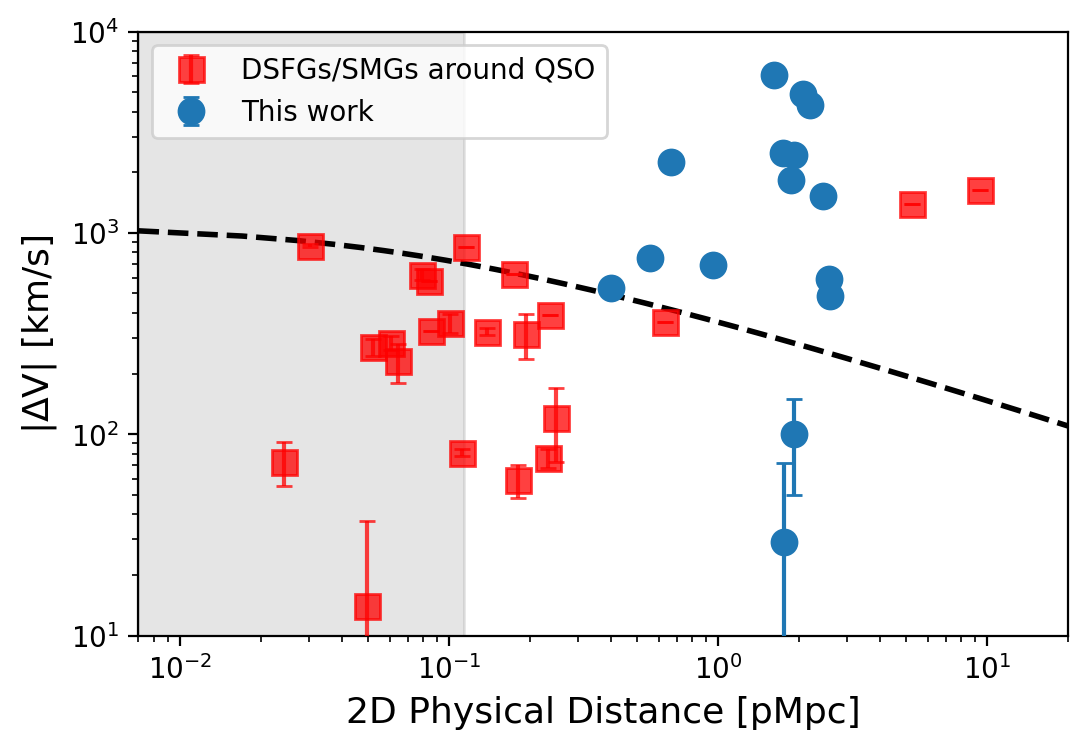}
\caption{Relative velocity to the host quasars vs the projected distance. Red points represent the DSFGs/SMGs around QSO \citep{2022ApJ...929..159C,2022ApJ...930...72A, Garcia2022, Li2023, 2024A&A...684A.119P, 2024arXiv240616637W}. Blue points show our sample SMGs. The shadowed region highlights the range of $R_{vir}$ for a halo of $10^{12.5}$\,$M_\odot$ at $z=3$. The curve represents the escape velocity for halos with masses of $10^{12.5}$ $M_\odot$ at $z \sim 3$.
\label{fig:fig10}}
\end{figure}

To study the dynamical state of the DSFGs/SMGs, we calculate the escape velocity of quasar halos assuming an NFW halo potential (\citealp{1996ApJ...462..563N})
\begin{displaymath}
\phi(s) = -g(c)\frac{ln(1+cs)}{s} V_{vir}^2
\end{displaymath}
where $g(c)=(ln(1+c)-c/(1+c))^{-1}$, $c$ is the concentration ($c = 3.5$; \citealt{2018MNRAS.479.3879W}; \citealp{2014MNRAS.441.3359D}), and $s = r/R_{vir}$. In \autoref{fig:fig10}, we plot the distribution between relative velocity and distances between DSFGs and their corresponding quasars, as well as the escape velocity (curve in \autoref{fig:fig10}) for halo masses of $10^{12.5}$ $M_\odot$ at $z\sim3$. 
The results suggest that DSFGs at projected distances smaller or comparable with 
the virial radius are gravitationally bound given the quasar halo mass of $\sim10^{12.5}$\,$M_\odot$. On the other hand, our sample SMGs outside the virial radius are consistent with 
being not bound. Given the large distance separations, they are most likely part of the Hubble flow.  


 In summary, the DSFGs that are located inside the quasar halo (hundred kpc scale) have a lower gas fraction, which is close to that of the central quasar. The relative velocity is well-constrained by the escape velocity of the halo potential. While the gas fractions of the lower luminosity sources ouside the quasar halo remain to be measured, the current comparison suggests that DSFGs located outside the central quasar halo range (Mpc scale) may have higher gas fractions comparable to those of field sources.
These results suggest that the surrounding DSFGs
 start to be impacted by the quasars environment when approaching the quasar's halo virial radius.
 The physical causes for the gas depletion could be gas stripping or strangulation, as recently shown by a detailed study of a DSFG around the Slug ELAN \citep{2021ApJ...923..200C}, consistent with some recent predictions from semi-analytic galaxy evolution models \citep{Ayromlou:2021aa}. 

\begin{figure}[h!]
\centering
\includegraphics[width=8.5cm]{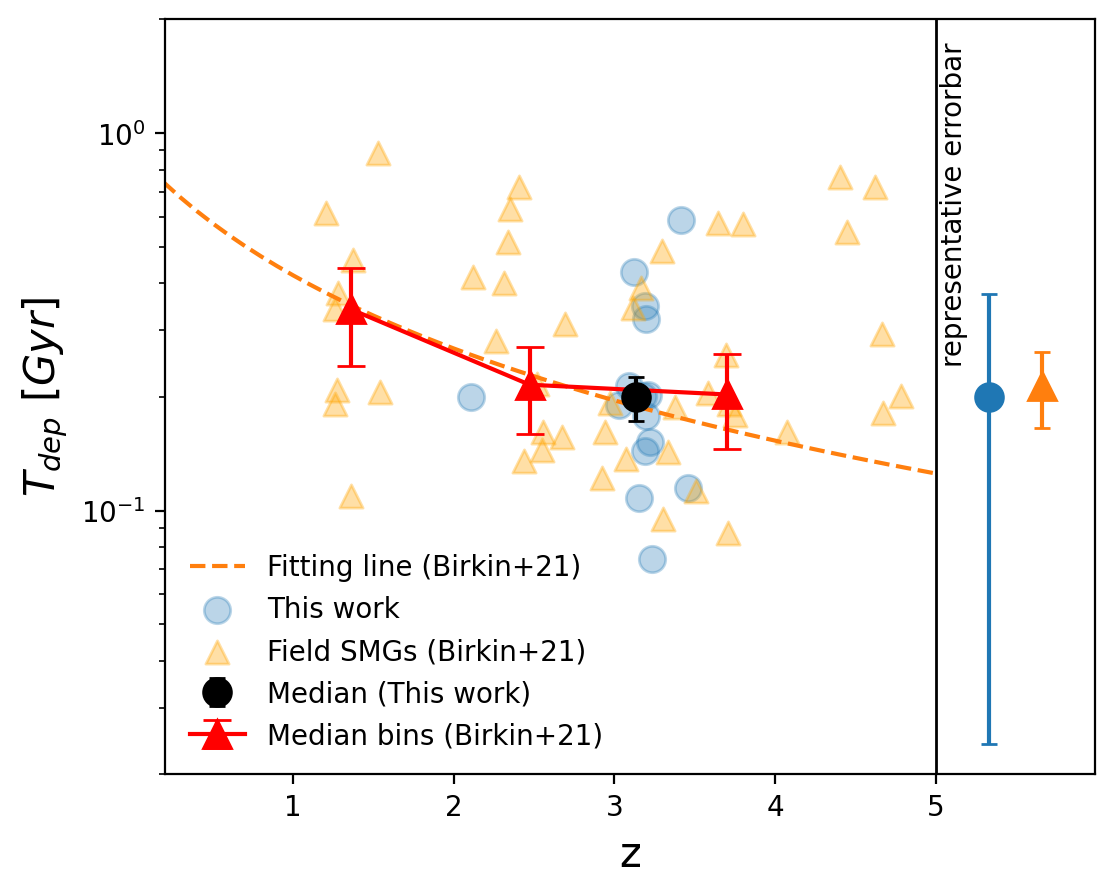}
\caption{Depletion timescales at different redshifts. The blue points show our primary samples from ALMA and NOEMA observation at $z\sim2-3$, with the black point and error bar representing the median value and the bootstrapped uncertainty (0.20\,$\pm$\,0.03 [Gyr]). Additionally, the orange triangles represent samples of field SMGs from \citet{2021MNRAS.501.3926B}, with the red  triangles and error bars indicating the median values. The orange dotted line is the fitting result of field SMGs provided by \citet{2021MNRAS.501.3926B}. The representative uncertainties for both this work and the field samples, calculated as the median uncertainty among individual sources from both groups, are shown in the right panel. }
\label{fig:fig11}
\end{figure}

\begin{figure*}[ht!]
\centering
\includegraphics[width=0.95\textwidth]{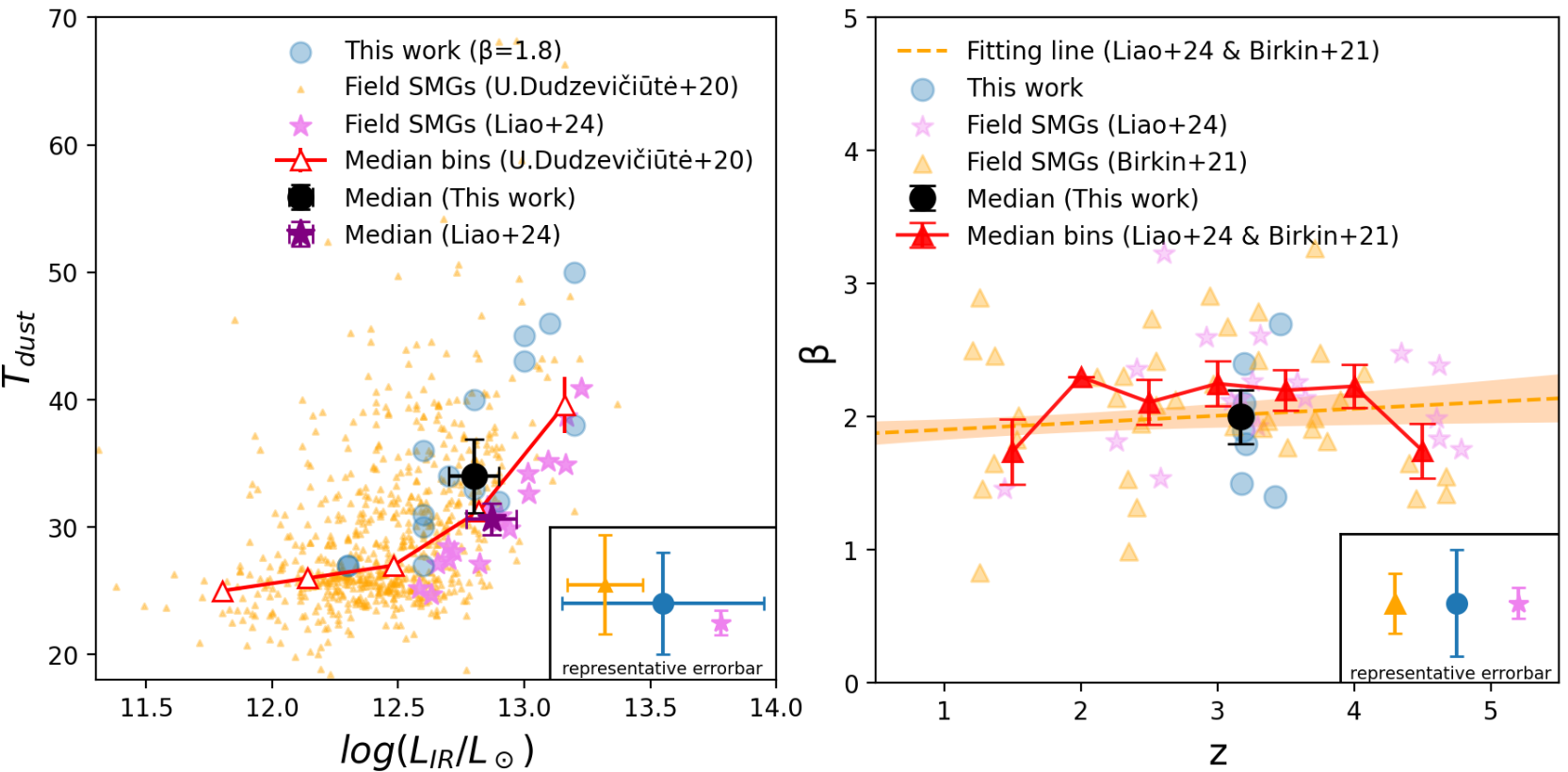}
\caption{\textbf{Left panel:} Dust temperature versus total infrared luminosity. The blue points show individual samples from this work. The purple points show AS2COSPEC SMGs presented in \citet{2024ApJ...961..226L}, and the orange triangles represent typical field SMGs samples from \citet{2020MNRAS.494.3828D}. Additionally, the black solid datapoint shows our median value at $z\sim2-3$ with bootstrapped uncertainties ($T_{dust} = 34.0\,\pm\,2.9 \,[K]$), and the purple solid star shows the median from \citealp{2024ApJ...961..226L}, and the red hollow triangles represent the median value for various luminosity bins of field SMGs based on \citet{2020MNRAS.494.3828D}.
\textbf{Right panel:} Dust emissivity index versus redshift. The blue points show individual samples from this work. Our median value ($\beta = 2.0\pm0.2$) is shown as a black point. The field SMG samples are shown as purple stars and orange triangles, from \citet{2021MNRAS.501.3926B} and \citet{2024ApJ...961..226L}, respectively. The median values in various redshift bins of all field SMGs are shown as red triangles. The orange dotted line shows the fitting result presented in \citet{2024ApJ...961..226L}, indicating there is no evolution for $\beta$. The representative uncertainties for each group, calculated as the median uncertainty among individual sources from the group, are shown in the lower right corners of both the left and right panels.
\label{fig:fig12}}
\end{figure*}

\subsubsection{Depletion timescale}
Next, we investigate the presence of any difference in the depletion times of our sample of SMGs in comparison to those of known SMGs in the field. 
The depletion timescale is given by
\begin{displaymath}
T_{\rm dep} = \frac{M_{\rm gas}}{\rm SFR},
\end{displaymath}
which displays the timescale for the gas reservoir in SMGs to be entirely consumed by star formation, assuming that there is no gas inflow or outflow \citep{2020ARA&A..58..157T}. On the other hand, the depletion timescale is the inverse of the star-formation efficiency. To derive this property, we calculate $M_{gas}$ by adopting $\alpha_{\rm CO} = 1.0$, same as the value adopted by studies of field SMGs \citep{2021MNRAS.501.3926B}. The SFRs are based on the total infrared luminosities derived in Section \ref{sec:sed} and listed in Table~\ref{tab:tab6}, and calculated by
\begin{displaymath}
{\rm log(SFR\,[M_{\odot}\,yr^{-1}]) = log(}\rm L_{IR}\,[\rm erg/s])-43.41
\end{displaymath}
, given by \citet{Kennicut2012}.

In general, the depletion timescale primarily depends on redshift and the offset from the main sequence (\citealp{2020ARA&A..58..157T}). Here, we examine the relationship between depletion timescale and redshift for our primary sample SMGs and the field SMGs in Figure \ref{fig:fig11}. For our sample at $z\sim3$, we calculate a median depletion timescale of 0.20\,$\pm$\,0.03 [Gyr], which is consistent with the trend of field SMGs (median of 0.17\,$\pm$\,0.05 [Gyr] at similar redshift) reported by \citet{2021MNRAS.501.3926B} within uncertainties. we therefore find no significant difference in depletion time between our SMGs in the dense environments and those in the field.

\subsubsection{Dust properties}
We also investigated the dust properties of our sample SMGs. In this section, we mainly focus on two properties, $T_{\rm dust}$ and $\beta$, which we derive from MBB fitting as usually done in the literature \citep{2020MNRAS.494.3828D, 2021MNRAS.501.3926B, 2024ApJ...961..226L}, and adopt the optically-thin model. If the general model is adopted instead, it would not significantly alter the value of $\beta$, as the Rayleigh-Jeans tail of the SED is well constrained by the long-wavelength data from ALMA and NOEMA. However, it would systematically yield a higher $T_{\rm dust}$ due to differences in model assumptions---specifically, the general model accounts for optical depth effects, while the optically thin model does not
\citep{2024ApJ...961..226L, 2021ApJ...919...30D}. To make fair comparisons, we adopt the same assumptions as those employed by previous studies.

First, in the left panel of Figure~\ref{fig:fig12}, we plot our $T_{\rm dust}$-$L_{\rm IR}$ distribution alongside those of typical field SMGs \citep{2020MNRAS.494.3828D,2024ApJ...961..226L}. 
Because \citet{2020MNRAS.494.3828D} adopted a fixed $\beta$ = 1.8, to make a fair comparison, the results plotted in this figure are based on the same assumption of $\beta$. The correlation of $T_{\rm dust}$-$L_{\rm IR}$ may result from sample selection effects. For example, the positive trend of $T_{\rm dust}$-$L_{\rm IR}$ for the SMGs is likely due to their selection at 850/870\,$\mu$m, which tends to bias the sample toward lower $T_{\rm dust}$ for sources with fainter $L_{\rm IR}$. This is the primary reason why we compare only to field SMGs instead of DSFG samples selected in other wavelengths. As shown in the figure, our sample SMGs follow a trend consistent with that of the field SMGs.

Next, we focus on the dust emissivity index, $\beta$, which impacts the SED in the Rayleigh-Jeans tail (low frequency side).  Initially, we assumed a fixed $\beta$ = 2.0 for the MBB fitting. To verify this assumption, we performed free-$\beta$ fitting again for 8 SMGs in our sample
that have robust Herschel detections (results are shown in Table~\ref{tab:tab6}), and derived a median value of $2.0 \,\pm\, 0.2$. This is consistent with the general assumption of $\beta$ = 2.0. We plot our median value in the distribution between $\beta$ and redshift (right panel of Figure~\ref{fig:fig12}). According to \citealp{2024ApJ...961..226L}, there is no significant evolution of $\beta$, and our median value agrees with this indication.

In summary, the dust properties ($T_{dust}$ and $\beta$, with median values of $34.0 \,\pm\, 2.9$ [K], $2.0 \,\pm\, 0.2$, respectively) of our SMG samples do not differ significantly from those of SMGs in the field (median values of $31.0 \,\pm\, 0.6$\,K and $2.2 \,\pm\, 0.2$ for $T_{\rm dust}$ and $\beta$, respectively, at similar $L_{\rm IR}$ and redshift).

\subsubsection{Gas to dust ratio}
Finally, we compare the gas-to-dust ratio, $\delta_{gdr}$, between our sample of SMGs and typical field SMGs, which implies the relationship between the molecular gas and dust content, defined as
\begin{displaymath}
\delta_{gdr} = \frac{M_{gas}}{M_{dust}}
\end{displaymath}

\begin{figure}[h!]
\centering
\includegraphics[width=8.5cm]{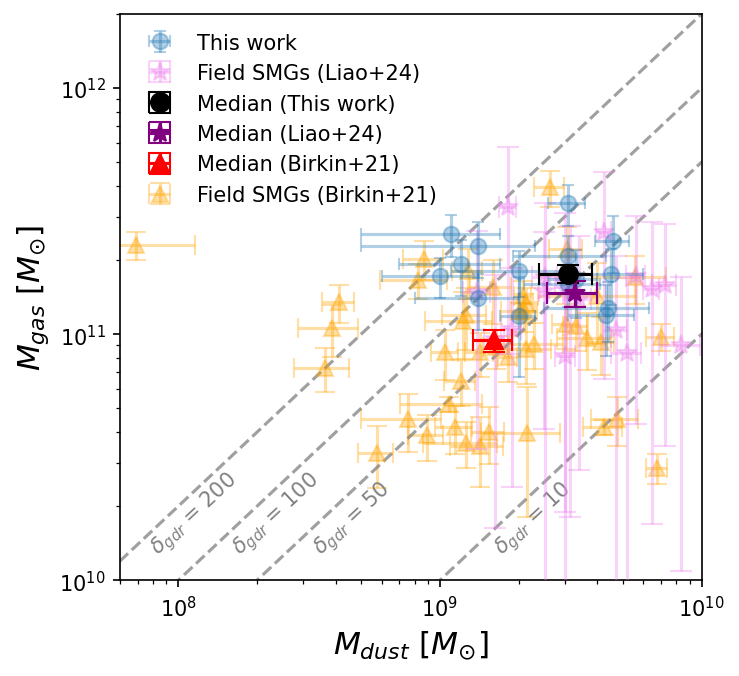}
\caption{Gas-to-dust mass distribution. The blue points represent our $z\sim2-3$ SMGs around quasars, with a median value of gas-to-dust mass ratio of 57 $\pm$ 14 (black point). The orange triangles represent the field SMGs from \citealp{2021MNRAS.501.3926B}, with a median of 59 $\pm$ 12 (red triangle). The purple stars represent the brightest field SMGs from AS2COSPEC \citep{2024ApJ...961..226L}, with a median of 44 $\pm$ 11 (purple star). The diagonal lines represent various gas-to-dust mass ratios, with the corresponding values indicated next to the lines.
\label{fig:fig13}}
\end{figure}

Generally, this property depends on metallicity, with more massive galaxies having lower $\delta_{gdr}$ due to a higher dust content \citep{2019MNRAS.490.1425L}. To compare this property, we show the gas mass versus dust mass distribution in Figure~\ref{fig:fig13}, with various $\delta_{gdr}$ values plotted as the diagonal lines in the figure. We plot our sample and field SMGs, focusing on the median value of each sample. Note that the dust masses used here for both our sample SMGs and field SMG are calculated based on the same MBB fitting assumptions. Our final result shows that our sample SMGs are consistent with typical field SMGs \citep{2021MNRAS.501.3926B,2024ApJ...961..226L} also for this parameter (the median value for each sample can be found in Figure~\ref{fig:fig13}).

\subsubsection{Summary on gas and dust properties}
Our results in this section point to a general picture in which the gas and dust properties of SMGs located beyond the virial radii of quasar halos are not significantly affected by their presumably denser surroundings. For DSFGs/SMGs that are within the halo, we observe evidence of decreasing gas fractions. Our results could help reconcile some discrepancies in previous studies, where the lack of observed environmental dependence of gas properties with environment could be due to the fact that the target sources were located too far from the host quasar or radio galaxy (e.g., \citealt{2017ApJ...842...55L}), and the impact could be easier detected when the studies focus on the core where the presence of a hot halo may lead to ram pressure stripping or strangulation (e.g., \citealt{2021ApJ...923..200C}), or probability of galaxy interactions increases (e.g., \citealt{Hughes:2025aa}).

\subsection{Spatial Distribution}
Cosmological simulations of structure formation predict that galaxies are embedded in a "cosmic web", which is composed of nodes of highest densities and numerous Mpc-scale substructures called "filaments" \citep{2014MNRAS.441.2923C, 2024A&A...684A..63G}. These filaments typical have lengths of $\sim1-100$\,cMpc \citep{Galarraga-Espinosa:2024aa}. Since our observations probe similar scales, in this section, we discuss the spatial distributions of our sample SMGs around quasars, which given the averaged halo masses of SMGs and quasars, could trace the filaments or nodes of the cosmic web.

Indeed, in the original SCUBA-2 survey \citep{FAB2023}, from which our sample is drawn, the sky distribution of SMGs around quasars reveals an elongated structure, reminiscent of a large-scale filament, after the fields were rotated by an angle $\alpha$ onto a common axis and stacked. The rotation angle represents the direction of 850~$\mu$m overdensities or the major axis of the nebular Ly$\alpha$ emission around each quasar. While that result is a promising indication of SMGs and quasars tracing large-scale filaments, the analyses were based on two dimensional distributions. As we now have confirmed members with redshifts, we can also investigate the distribution in the third dimension. 

Due to the small sample size of SMGs with spectroscopic redshifts, we can only perform statistical analyses; We first plotted the sky locations of SMGs centered at the quasars in the targeted fields. 
The result is shown in the left panels in Figure \ref{fig:fig14}. We have redshift information for every SMGs in our samples, which allows us to derive the relative distance from SMGs to central quasars along the z-axis (light-of-sight direction). Therefore, we also plotted the x(y)-z plane distribution. Note that since our SMGs are located Mpc away from the central quasars so sitting in the Hubble flow (\autoref{fig:fig9}), the redshift differences likely indicate differences in radial distance, rather than peculiar velocity.

\begin{figure}[ht!]
\centering
\includegraphics[width=0.47\textwidth]{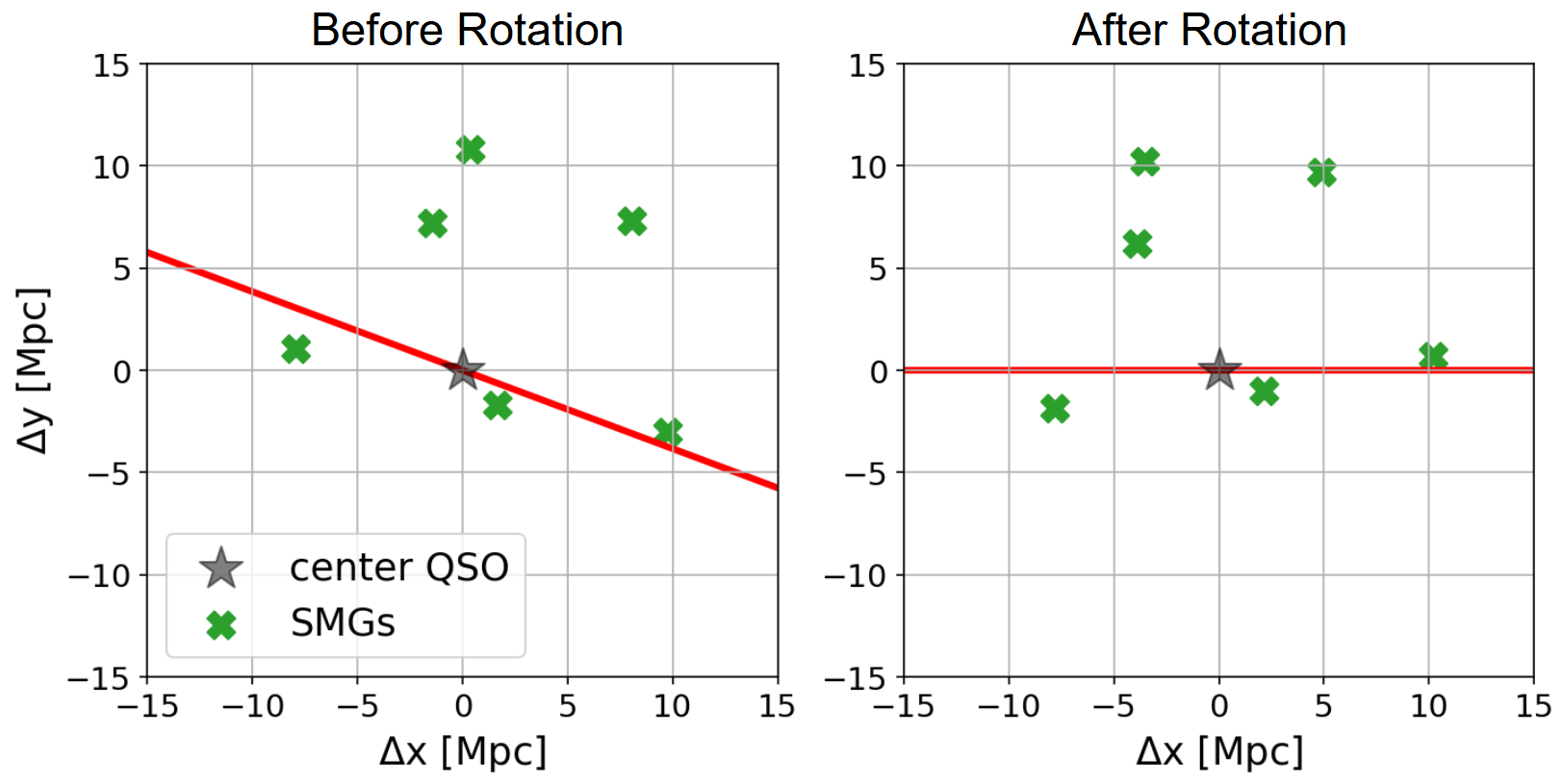}
\caption{ An example for axis rotation based on the field SDSSJ0819. The green crosses represent the relative locations for primary samples to the central quasar. The red line shows the direction of filament reported by \citet{FAB2023}. The left panel show the real distribution, whereas the right panel shows the case after rotating the field such that the direction of filament is aligned with the x-axis.
\label{fig:rot}}
\end{figure}

At first glance, there is no coherent structure in the x-z plane. However, even if there are large-scale structures traced by SMGs around the targeted quasars, 
we would expect a fully random distribution if we simply stack all maps for each targeted field. 
\begin{figure*}[ht!]
\centering
\includegraphics[width=\textwidth]{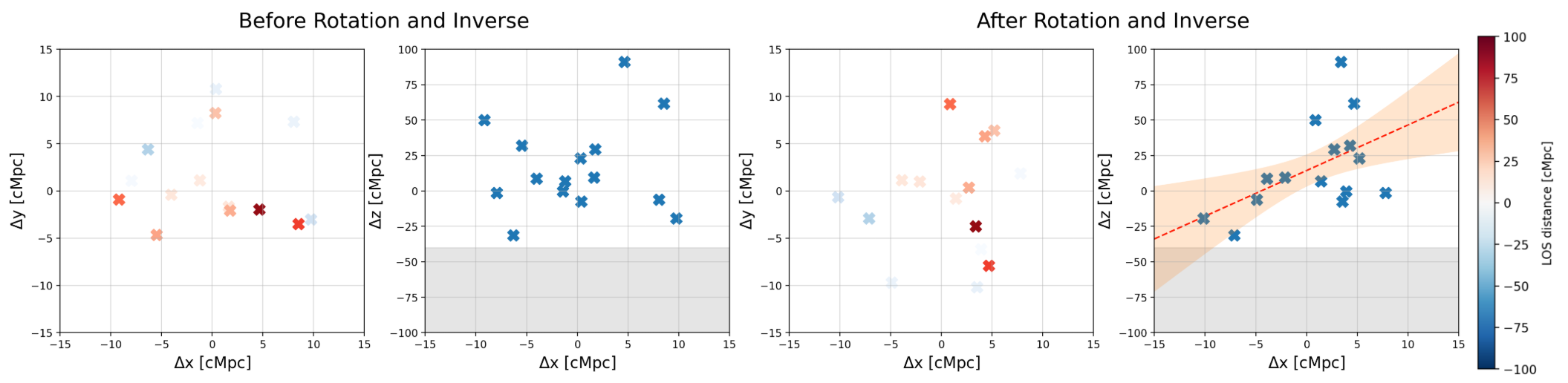}
\caption{The left two panels show the stacked spatial distribution (x-y, x-z) in the original situation (before rotation), while the right two panels show the stacked distribution (x-y, x-z) after rotation for each quasar field (using the angles from Table 6 in \citealp{FAB2023}), with SDSSJ0819, SDSSJ1020 and SDSSJ1209 inverted (best realization with lowest p-value). The color code in the x-y distribution represents the comoving distance along the line-of-sight direction. Blue indicates closer distances and red indicates farther distances, with the color bar shown on the right side of the figure. The red dotted lines in the x-z distribution represent the fitting results with 1-$\sigma$ uncertainty derived from the linear regression, with the uncertainty given by the standard deviation. The shaded region in the x-z distribution shows the area where sources cannot be detected due to the limited spectral window coverage of the data.
\label{fig:fig14}}
\end{figure*}

We therefore adopted the rotation angles of the filament on the plane-of-sky, $\alpha$, derived from 2D spatial distribution reported by \citet{FAB2023} and then rotated each quasar field to ensure that all filaments are aligned in the same direction. An example is shown in Figure~\ref{fig:rot}. Importantly, the filament spine is assumed to be a straight line and the angles $\alpha$ are derived from the minimum sum of distances from each SMG to the x-axis after rotation, such that the filament will lie on the x-axis. For this reason, the direction of the filament as described by $\alpha$ would be the 
same 
if rotated by $180^{\circ}$.
\par
Because of this degeneracy of distribution caused by the rotation angle, we proceed as follows: We consider the redshift information now available for the SMGs and allow for field inversion (rotate by $180^{\circ}$). If the detected SMGs around the quasars trace a large-scale structure filament, we expect to find an alignment 
in the x-z plane distribution (since the filament lies on the x-axis) and observe a strong correlation in the distribution. To find the strongest correlation in the x-z distribution, we tested every possible inversion scenario for 
the quasar field studied and applied Pearson correlation testing to each scenario. We finally found the lowest p-value = 0.05 in the scenario where SDSSJ0819, SDSSJ1020 and SDSSJ1209 were inverted. The result after rotation and inversion is shown in the right panels of \autoref{fig:fig14}. 

\begin{figure}[ht!]
\centering
\includegraphics[width=8.5cm]{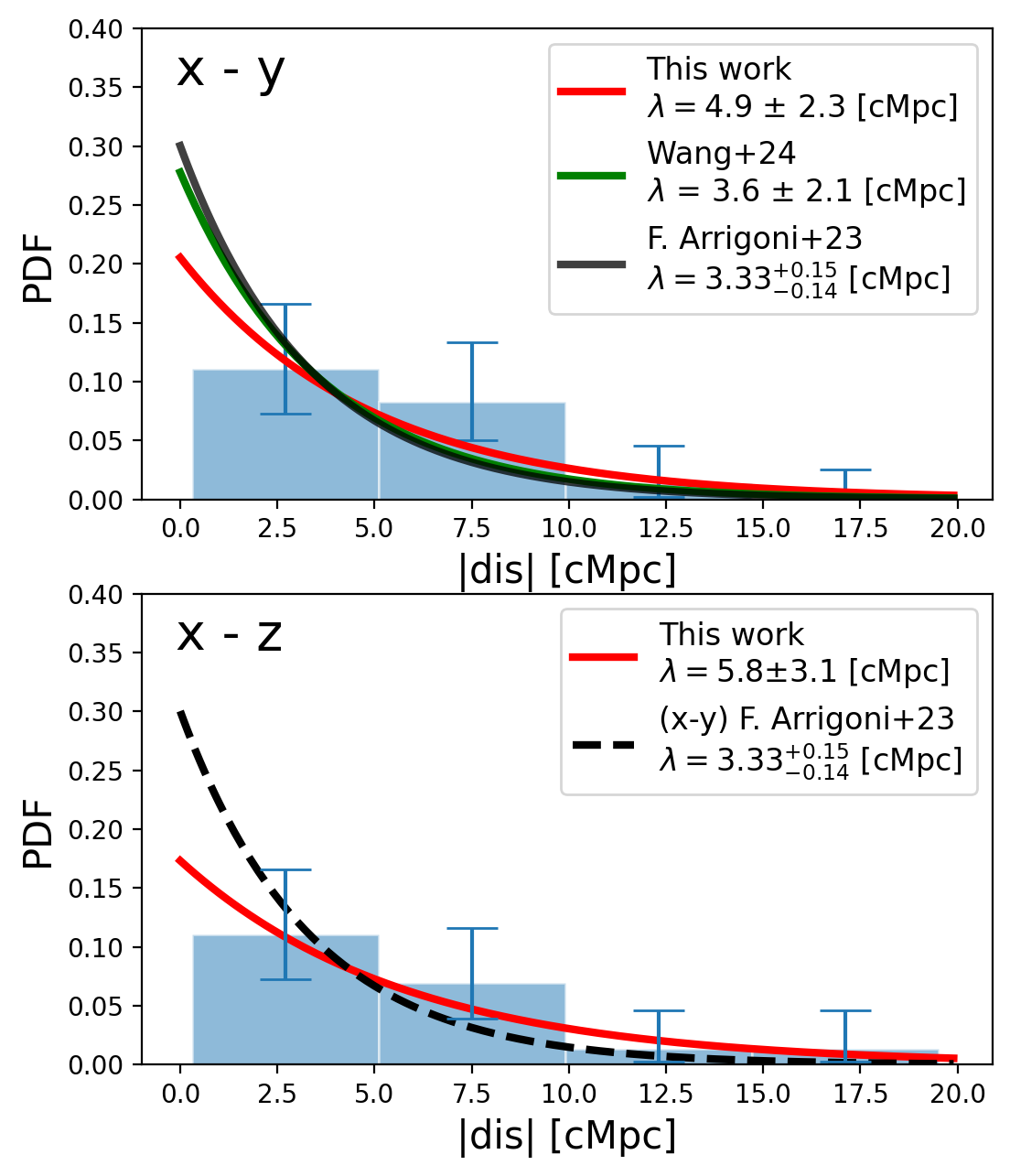}
\caption{Probability distributions of projected distances ($|dis|$) between the SMGs and the probable large-scale spine structures, in both the projected sky plane (x-y; upper panel) and the projected line-of-sight plane (x-z; lower panel). 
The error bars are calculated from Poisson error.  Red curves represent the fitting results of our sources for the filament widths, green curve represents the outcome after applying our methods to sources reported in \cite{2024arXiv240616637W}, and black curves represent the x-y distribution results from \cite{FAB2023}. 
\label{fig:fig15}}
\end{figure}

The next thing we are interested in is the width of the structure or filament. According to \citealp{FAB2023}, the empirical probability distribution (ePDF) of $|y|$ (distance to the x-axis) of each SMG in the quasar fields is related to the width of the filament by this formula:
\begin{displaymath}
PDF(|y|) = exp(-|y|/\lambda)/\lambda,
\end{displaymath}
where $\lambda$ is the scale width of the filament. We then apply this to our x-y and x-z distributions, noting that in x-z distribution, we focus on the PDF of $|dis|$: the distance from SMGs to the best fit line 
($\Delta z = 3.2\Delta x+14.4$ [cMpc]). The PDFs of $|y|$ and $|dis|$ are shown in the upper and lower panels of \autoref{fig:fig15}, respectively. After applying the same formula to estimate the width of the filament in both directions, we derive $\lambda = 4.0 \pm 2.6$ cMpc in the x-y plane and $\lambda = 3.5 \pm 2.2 $ cMpc in the x-z plane. Although there is a large uncertainty, both results are consistent within uncertainties with the results reported by \citet{FAB2023}. 
We have also applied our method to another sample of 18 SMGs that are spectroscopically confirmed to be associated with a $z = 2.85$ proto-cluster hosted by the quasar HS~1549+19 \citep{2024arXiv240616637W}, and obtained $\lambda = 3.6 \pm 2.1 $ cMpc in the sky projected (x-y) direction, which is in good agreement with our results.
According to the results above, because the widths of the filament are consistent with each other in both the x-y and x-z directions, it suggests that the large-scale filaments traced by quasars and DSFGs resemble the shape of a cylinder or elongated pancake.

In \autoref{fig:fig15}, it can be seen that the maximum values of $|dis|$ in the x-y and x-z distributions are different. This is because our sample is a sub-sample of the sources used in \citet{FAB2023}, which is limited to the footprint of SCUBA-2 observations, roughly a circular area with a radius of $\sim$ 10 cMpc. This limitation imposes a boundary on the PDF in the x-y distribution. On the other hand, 
in the x-z distribution because $|dis|$ is determined not only by the x-axis value but also by the z-axis value, which is limited by the wavelength range of the spectral window ($\sim$-40-100 cMpc).

To estimate the statistical significance of this detected filament structure, we performed a simulation to derive the probability of reproducing the observed distribution and the filament width. First, we fixed the locations in the x-y distribution to match the map after rotation and inversion. Next, random values were assigned to the z-axis while keeping the blue and red fixed, i.e., fixing the positive and negative values of the z-axis, mimicking the inversion processes done on the real data. We repeated this process 1000 times and performed Spearman correlation testing for each condition. We found that the ratio of the p-values $\leq$ 0.05 (same as our result) is 0.5\% (significance between 2-$\sigma$ and 3-$\sigma$). We then applied the same method to deduce the width of the filament in the mock filaments with p-values $\leq$ 0.05, and found them consistent with $\lambda = 3.33$ (\citealp{FAB2023}) within the uncertainties. This simulation indicates that there is about a 0.5\% probability that our observed result is consistent with random distribution, which means that our reported detection of filaments is tentative. To increase the statistical significance of our results and decrease the uncertainty.
We tested this by including the supplementary sources\footnote{These sources differ from our primary sample as they do not have 850\,$\mu m$ detection.} that fall within the velocity range associated with the proto-cluster ($\pm 7000\, \rm km/s$) in our analysis. We found a marginal improvement in the significance of the result.

\subsection{Cumulative SFR}
Measurements of total SFRs and their spatial distributions in protoclusters can help shed light on the physical mechanisms of mass assembly in some of the densest structures in the Universe, and in turn put strong constraints on state-of-the-art galaxy cluster formation models \citep{Bassini:2020aa,Lim:2021aa,2022MNRAS.509.4037Y,Remus:2023aa,Araya-Araya:2024aa}. In this section, 
we calculated the cumulative SFR within the survey area at $z\sim2-3$. In addition, we computed the total SFR per unit comoving volume to estimate the star formation rate density (SFRD). We also compare our results with those reported by other observations and simulations in the literature.

\begin{figure*}[ht!]
\centering
\includegraphics[width=\textwidth]{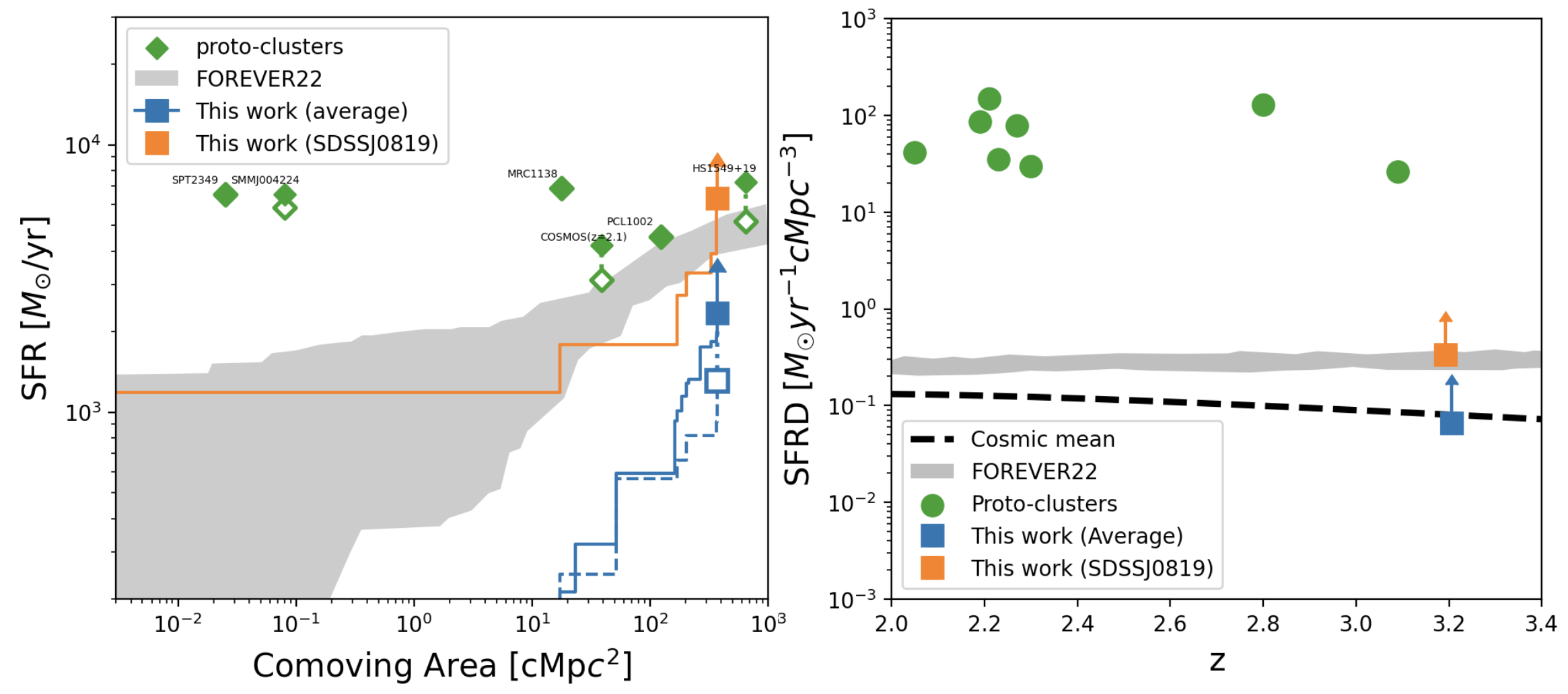}
\caption{\textbf{Left panel:} Cumulative star-formation rate. Our measurements for SDSSJ0819 and for all of our quasar fields are shown as orange and blue squares, respectively. Since only bright sources are targeted by our observations these measurements are likely lower limits. The green solid diamonds show the observed cumulative SFRs of other proto-cluster candidates: PCL1002 \citep{2015ApJ...808L..33C}, COSMOS(z=2.1)
\citep{ 2016ApJ...826..130H}, SMMJ004224 \citep{2018ApJ...856...72O}, MRC1138-256 \citep{2014A&A...570A..55D}, HS1549+19 \citep{2024arXiv240616637W}, and SPT2349-56 \citep{2018Natur.556..469M}. The hollow symbols represent the values derived from the corresponding samples but limited to sources located within the light-of-sight depth of 28.6\,cMpc, as defined by the FOREVER22 simulation. 
The gray region represents the predicted range from all simulated halos with masses comparable to those of quasars in the FOREVER22 simulation \citep{2022MNRAS.509.4037Y}.
\textbf{Right panel:} Star-formation rate density as a function of redshift. 
The orange and blue solid squares are the results of calculations based on the confirmed member galaxies that are located within the volume centered at quasars with a size specified by the FOREVER22 simulation, (25.7 cMpc$)^3$, for SDSSJ0819 and the average for all of our quasar fields, respectively. Again, they are likely lower limits. The green points show some of the measurements from the literature \citep{Clements2014, kato2016}. The gray region represents the predicted region based on all simulated halos with masses comparable to those of quasars in the FOREVER22 simulation. Furthermore, the cosmic mean curve reported by \citet{Madau2014} is shown as a black dashed line.
\label{fig:fig16}}
\end{figure*}

In our data, we calculate the cumulative SFR and SFRD for SDSSJ0819 (the field with the most detections) and the average value for all fields in our observation. The cumulative SFRs for SDSSJ0819 and the field average are shown in the left panel of \autoref{fig:fig16}. Note that our cumulative SFRs should only be considered as lower limits. That is because our ALMA observations only targeted the top 10-20 brightest objects based on their S/N at 850\,$\mu m$, resulting in incompleteness. Without deeper surveys that allow us to uncover the true distribution of lower SFR sources, it is difficult to apply a reliable correction for incompleteness. The same holds for the calculations of the SFRDs, where our measurements are likely lower limits, as shown in the right panel of \autoref{fig:fig16}. 

To make comparisons with results in the literature, we first make use of a cosmological hydrodynamical simulation called FOREVER22 \citep{2022MNRAS.509.4037Y}. 
In FOREVER22, the \textsc{MUSIC} code \citep{Hahn2011} was used to generate the initial conditions. An $N$-body simulation was then performed using $256^3$ dark matter particles within a full simulation box of volume $(714.2~\mathrm{cMpc})^3$, designed to investigate the statistical properties of galaxies in protoclusters and their surrounding large-scale structures. The ten most massive haloes at $z = 2$ were selected, and zoom-in initial conditions were generated for each, with a side length of $28.6\,\rm cMpc$. Simulations were run from $z = 100$ down to $z = 0$, with a total of 200 snapshots uniformly spaced in time for each region. By $z = 0$, these regions had evolved into massive haloes with halo masses $M_{\mathrm{halo}} \geq 10^{14}~M_\odot$, which were found to reside at the nodes or intersections of large-scale filaments. Due to potential inaccuracies in hydrodynamic and gravitational calculations near the boundaries of the simulation boxes, certain statistical properties were analyzed only within the inner, defined by a volume of $(25.7~\mathrm{cMpc})^3$. Consequently, the simulation provides predictions for cumulative SFRs and total SFRDs around haloes with masses comparable to those of quasars ($\sim 10^{13}~M_\odot$) at $z = 3$. 
In addition, observational studies are also used for comparison with our results \citep{Clements2014,kato2016,2015ApJ...808L..33C, 2016ApJ...826..130H, 2018ApJ...856...72O, 2014A&A...570A..55D, 2024arXiv240616637W, 2018Natur.556..469M}. 
These theoretical and observational results are plotted in \autoref{fig:fig16}.

For cumulative SFRs, we see that model predictions generally agree with our measurements, as well as some measurements from the literature. However, since our measurements are likely lower limits, it is possible that with deeper observations, the true underlying values could exceed the model predictions, especially for SDSSJ0819. We see a similar behavior in the comparison on SFRDs, where the models generally agree with our measurements. 

Note that for calculations of both cumulative SFRs and SFRDs, to make a fair comparison with the model predictions, we adopt the volumes defined in the simulations, which means a projected depth of 28.6\,cMpc and a volume of (25.7 cMpc$)^3$, for cumulative SFRs and SFRDs, respectively \citep{2022MNRAS.509.4037Y}. As a result, we obtain SFRDs of 0.34 $\pm$ 0.07 $M_\odot\, \rm yr^{-1}\, cMpc^{-3}$ for SDSSJ0819, and 0.07 $\pm$ 0.04 $M_\odot\, \rm yr^{-1}\, cMpc^{-3}$ for all of our quasar fields, respectively. The match in volume was not considered in other observational works in the literature, which can potentially explain in part why the results of other works are much higher than the model predictions and our measurements. In fact, if we take the measurements reported in the literature and adopt the same volume for the calculations, some of the reported values for the cumulative SFRs become lower and closer to the model predictions (\autoref{fig:fig16}). We therefore highlight the need to clearly define volumes for calculations in order to make fair comparisons, especially with increasing bandwidths allowed by future surveys in studies of this kind.

\section{Conclusion} \label{sec:Conclusion}
To understand how the environment impacts dusty star-forming galaxies (DSFGs), we have targeted with ALMA and NOEMA the brightest 101 850\,$\mu$m SCUBA-2 detections ($\geq$ 4$\sigma$) in 9 $z\sim2-3$ quasar fields, 
which have been confirmed as regions with 850\,$\mu$m counts in excess by a factor of $\sim3-4$ with respect to blank fields
(\citealp{Nowotka2022,FAB2023}). 
We summarize the results in the following.

\begin{enumerate}
\item We obtained 28 line detections, of which 15 have additional detections in far-infrared or submillimeter continuum and spectroscopically confirmed to be physically associated with the quasar fields. The relatively low detection rates can be explained by jointly considering the original SCUBA-2 number counts, the incompleteness of the spectral coverage, and that about half of the physically associated sources lie below the current detection limit.

\item Using the 14 CO(4-3) and 1 CO(3-2) line emitters 
as the primary set for analyses, 
we find that most ($73^{+29}_{-21}\%$) of these detection are well-fit by a 
double Gaussian model, and the median peak-to-peak separation is $350 \pm 25$~km~s$^{-1}$, consistent with what has been reported by recent spectroscopic surveys of field SMGs \citep{2021MNRAS.501.3926B,2024ApJ...961..226L}. The relationship between line luminosity ($L'_{\rm CO}$) and line width is consistent with that found in field SMGs, in agreement with the findings of \citet{2017A&A...608A..48D}.

\item 
Together with the measured spectroscopic redshifts, we applied modified black-body (MBB) fitting to the far-infrared photometry on the 15 primary sources. Assuming a dust emissivity index $\beta$ = 2.0, we derived the median dust temperature $T_{dust}$ = 34 $\pm$ 3 K, consistent with the field SMGs at fixed infrared luminosity.
Additionally, we applied free-$\beta$ fitting for a few sources that have Herschel photometric measurements around the SED peak 
and obtained a median $\beta$ of 2.0 $\pm$ 0.2, which agrees with our assumption for the fixed $\beta$ fittings, as well as the recently reported values for the field SMGs \citep{2024ApJ...961..226L,Ward:2024aa}. In addition, in the comparison of depletion time and gas-to-dust ratio, the median values of our sample show a level comparable to that of field samples. This suggests that these dust and gas properties for SMGs around quasars on Mpc scales are consistent with those of SMGs residing in the field.


\item Using the relation between dynamical mass and gas mass \citep{2021ApJ...923..200C}, we estimate the gas mass fraction for both SMGs around quasars and in blank fields. Complementing our data with similar measurements of DSFGs around quasars and quasars themselves in the literature,
we were able to compare gas fractions of dusty galaxies
from quasar halo scales ($\sim100$~kpc) out to Mpc scales.
The results show that inside the quasar virial radius, DSFGs have lower gas fractions, at a level similar to that of the QSO host galaxies.
Outside the virial radius, DSFGs have a gas fraction comparable to the level of their field counterparts. 
However, due to the limited sensitivity of our data, fainter CO sources with luminosities comparable to those near quasars may remain undetected, and their gas fractions will require confirmation through deeper observations.
Based on the above results, it suggests that massive dusty galaxies are only affected by the quasar environment when they are approaching or within the halo scale, but the effect is unclear when they are outside of it.
\end{enumerate}

In summary, the points mentioned above reveal that dense environments significantly impact massive dusty star-forming galaxies (typical stellar mass of $\sim10^{11}$\,$M_\odot$; \citealt{2020MNRAS.494.3828D, 2024ApJ...961..226L}) only within the halo scales around massive halos ($\sim$$10^{12.5} M_{\odot}$) at cosmic noon. Furthermore, by focusing on the spatial distribution and cumulative star formation rate (SFR) of quasar fields, we conclude the following additional results.

\begin{enumerate}
\setcounter{enumi}{4}
\item 
Our statistical analyses allow us to obtain tentative detection on the coherent large-scale filament structures traced by dusty galaxies in quasar fields in both the projected sky (x-y) and projected radial (x-z) directions.
By estimating the scale widths of the filament, we obtained 4.9 $\pm$ 2.3 cMpc and 5.8 $\pm$ 3.1 cMpc for the x-y and x-z distribution, respectively. In addition to agreeing with recent results \citep{FAB2023, 2024arXiv240616637W}, the consistent filament width values for the two directional distributions indicate that the large-scale filament has a cylindrical or elongated pancake shape.

\item Considering the SFR of galaxies in the targeted quasar fields, 
we calculated the cumulative SFR and SFRD for SDSSJ0819 (the field with most confirmed sources) 
and the average for all the other fields. We find that our measurements are in overall good agreement with model predictions 
and observational results of other proto-cluster fields. However, since our observations only targeted the brightest sources, our measurements are likely lower limits. Deeper observations with more targets will help put better constraints on this issue.
\end{enumerate}

The points mentioned above provide first insights on the properties of SMGs in quasar fields and on the large-scale structures they trace. 
These results were achieved by focusing on a very small fraction (the brightest 17~\%) of the known $850$~$\mu$m sources in the targeted quasar fields and by using a very quick snapshot survey to allow for scheduling on the competitive ALMA and NOEMA arrays. 
Therefore, this work showcases the potential in targeting the plethora of 850~$\mu$m targets available in quasar fields and whose nature and association still remain unclear. 
To complete the picture of quasar environments -- not only the structure and properties of their field itself, but also to verify the differences between DSFGs/SMGs around quasars and in blank fields -- larger statistical follow-up observations with ALMA and NOEMA, and additional data at multiple wavelengths are needed.
\vspace{5mm}
\facilities{JCMT(SCUBA-2), ALMA(Band3 and 4), NOEMA(Band 1), Herschel(SPIRE)}
\software{\textsc{LineSeeker}(\citealp{2017A&A...597A..41G})
          }
          
\begin{acknowledgments}
We thank the reviewer for a careful report that improves the manuscript. We would also like to thank Yi-Kuan Chiang for useful discussions. We are grateful to the maintenance and administrative staff of our institutions, whose efforts in supporting our day-to-day work environment make our scientific discoveries possible. Y.-J.W. and C.-C.C. acknowledge support from the National Science and Technology Council of Taiwan (111-2112M-001-045-MY3), as well as Academia Sinica through the Career Development Award (AS-CDA-112-M02). R.D.\ acknowledges support from the INAF GO 2022 grant ``The birth of the giants: JWST sheds light on the build-up of quasars at cosmic dawn'' and the PRIN MUR ``2022935STW'', RFF M4.C2.1.1, CUP J53D23001570006 and C53D23000950006.
HD acknowledge support from the Agencia Estatal de Investigaci{\'o}n del Ministerio de Ciencia, Innovaci{\'o}n y Universidades (MCIU/AEI) under grant (Construcci{\'o}n de c{\'u}mulos de galaxias en formaci{\'o}n a trav{\'e}s de la formaci{\'o}n estelar oscurecida por el polvo) and the European Regional Development Fund (ERDF) with reference (PID2022-143243NB-I00/10.13039/501100011033).
This paper makes use of the following ALMA data: ADS/JAO.ALMA\#2019.1.01514.S. ALMA is a partnership of ESO (representing its member states), NSF (USA) and NINS (Japan), together with NRC (Canada), NSTC and ASIAA (Taiwan), and KASI (Republic of Korea), in cooperation with the Republic of Chile. The Joint ALMA Observatory is operated by ESO, AUI/NRAO and NAOJ.
This work is based on observations carried out under project number W18EM with the IRAM NOEMA Interferometer. IRAM is supported by INSU/CNRS (France), MPG (Germany) and IGN (Spain). The research leading to these results has received funding from the European Union’s Horizon 2020 research and innovation program under grant agreement No 101004719 [Opticon RadioNet Pilot ORP].”

\end{acknowledgments}

%




\appendix
\section{Overview for quasar field maps} \label{apx:a}
\begin{figure*}[h]
\centering
\includegraphics[width=15cm]{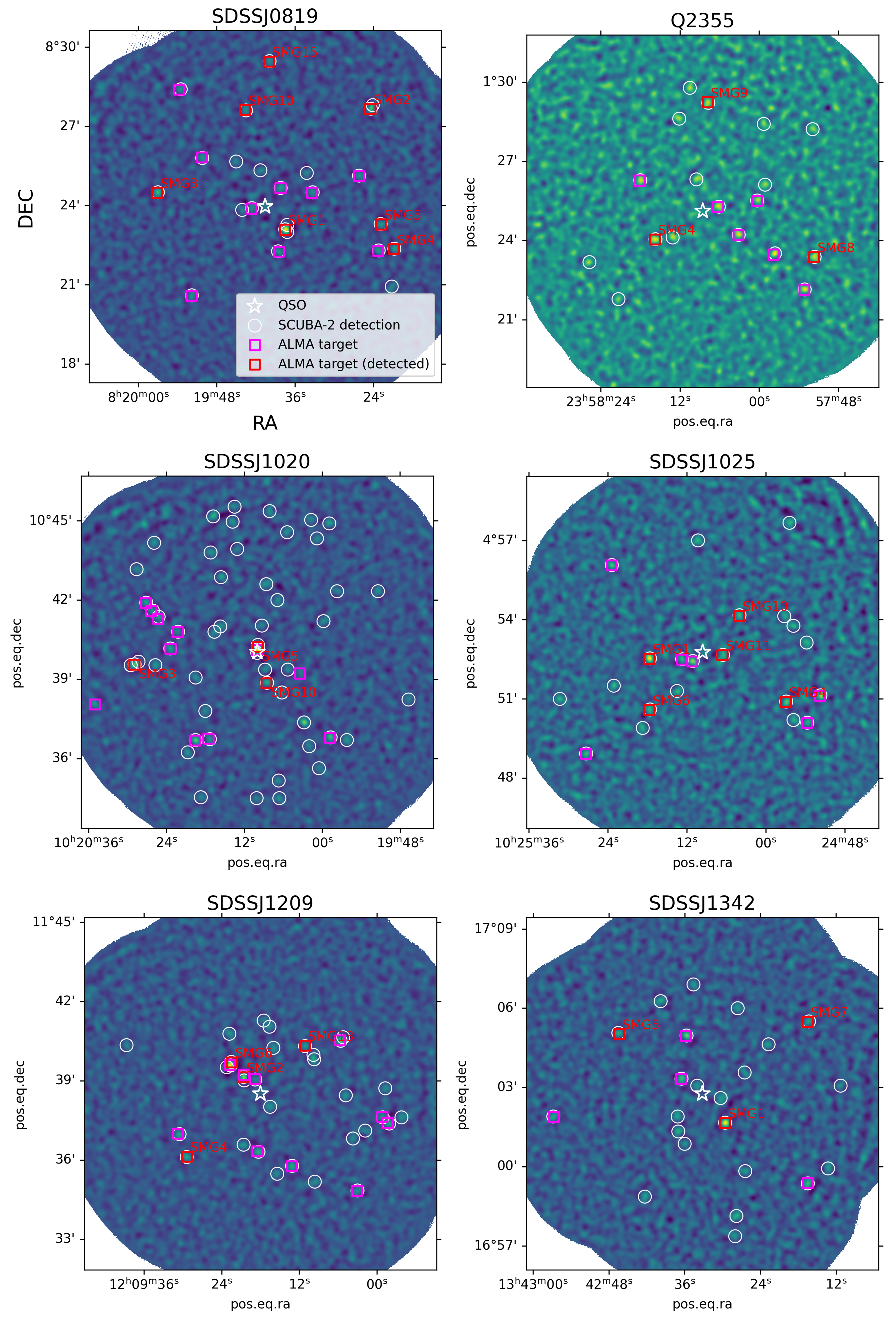} 
\caption{JCMT/SCUBA-2 850\,$\mu m$ SNR maps of quasar field hosting Ly$\alpha$ nebulae. The plotting style follows that of the upper panel in Figure~\ref{fig:fig.1}.
\label{fig:fig18}}
\end{figure*}

We list our 850\,$\mu m$ maps of quasar fields hosting Ly$\alpha$ nebulae in \autoref{fig:fig18}, in a same style as the upper left panel in \autoref{fig:fig.1}.

\section{CO emission maps} \label{apx:b}
\begin{figure*}[h]
\centering
\includegraphics[width=16cm]{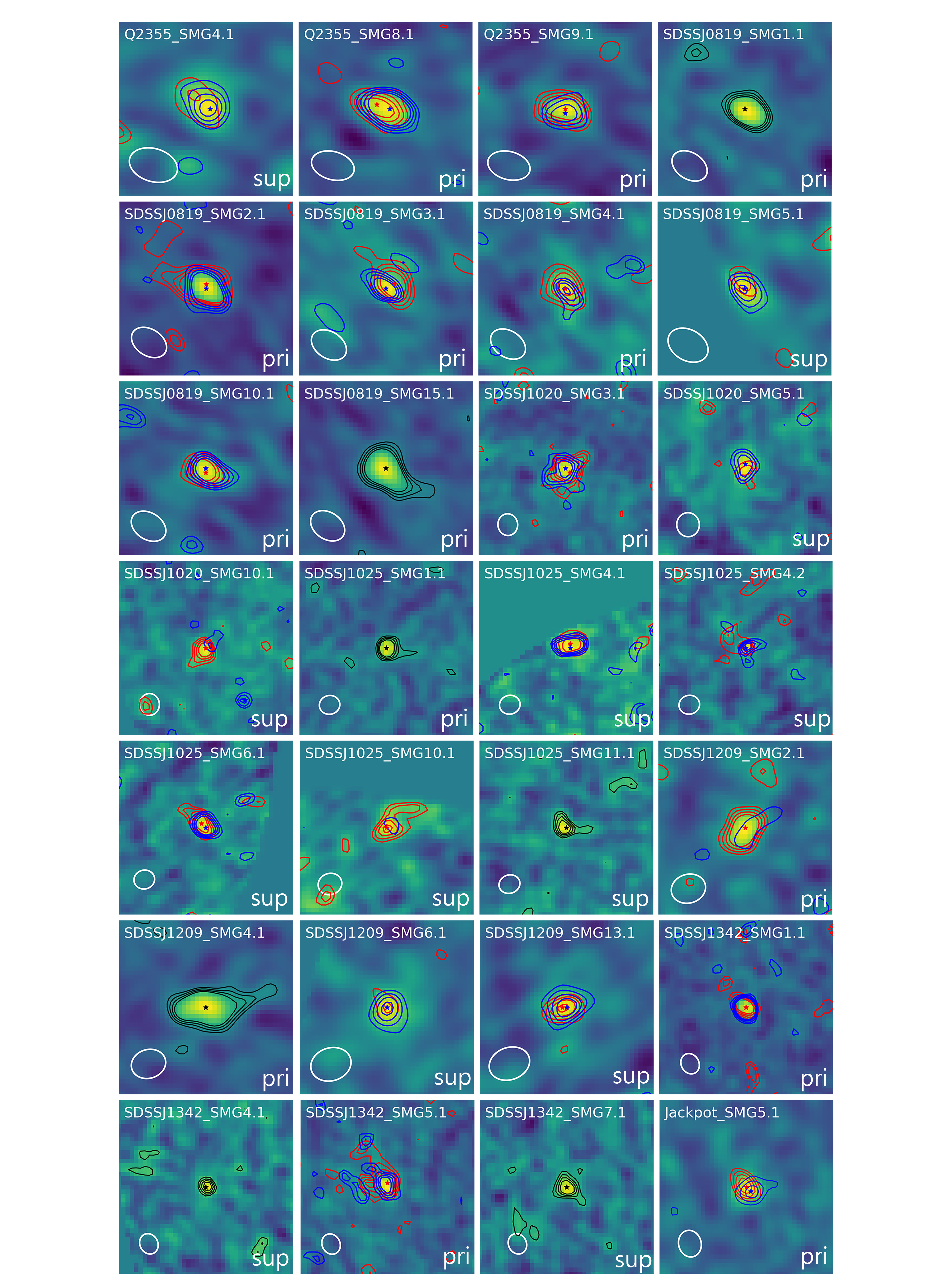}
\caption{CO emission maps for all the 28 primary and supplementary detections. The plotting style follows that of the lower panels of Figure~\ref{fig:fig.1}.
\label{fig:COmap}}
\end{figure*}

We show all the CO emitter maps stacked over the channels within the FWHM of the emission lines (\autoref{fig:COmap}). We perform CASA task \textsc{Imfit} to the emission maps. As expected, most CO emissions are unresolved. However, some sources (e.g., SDSSJ1020\_SMG3.1, SDSSJ1342\_SMG5.1) appear resolved. Follow-up observations with higher spatial resolutions are needed to confirm. 


\section{Spectrum for CO emission} \label{apx:c}
Here we show in \autoref{fig:spec2} all the spectra of the 28 detected line emitters from the ALMA/NOEMA cubes.
\begin{figure*}[h]
\centering
\includegraphics[width=18cm]{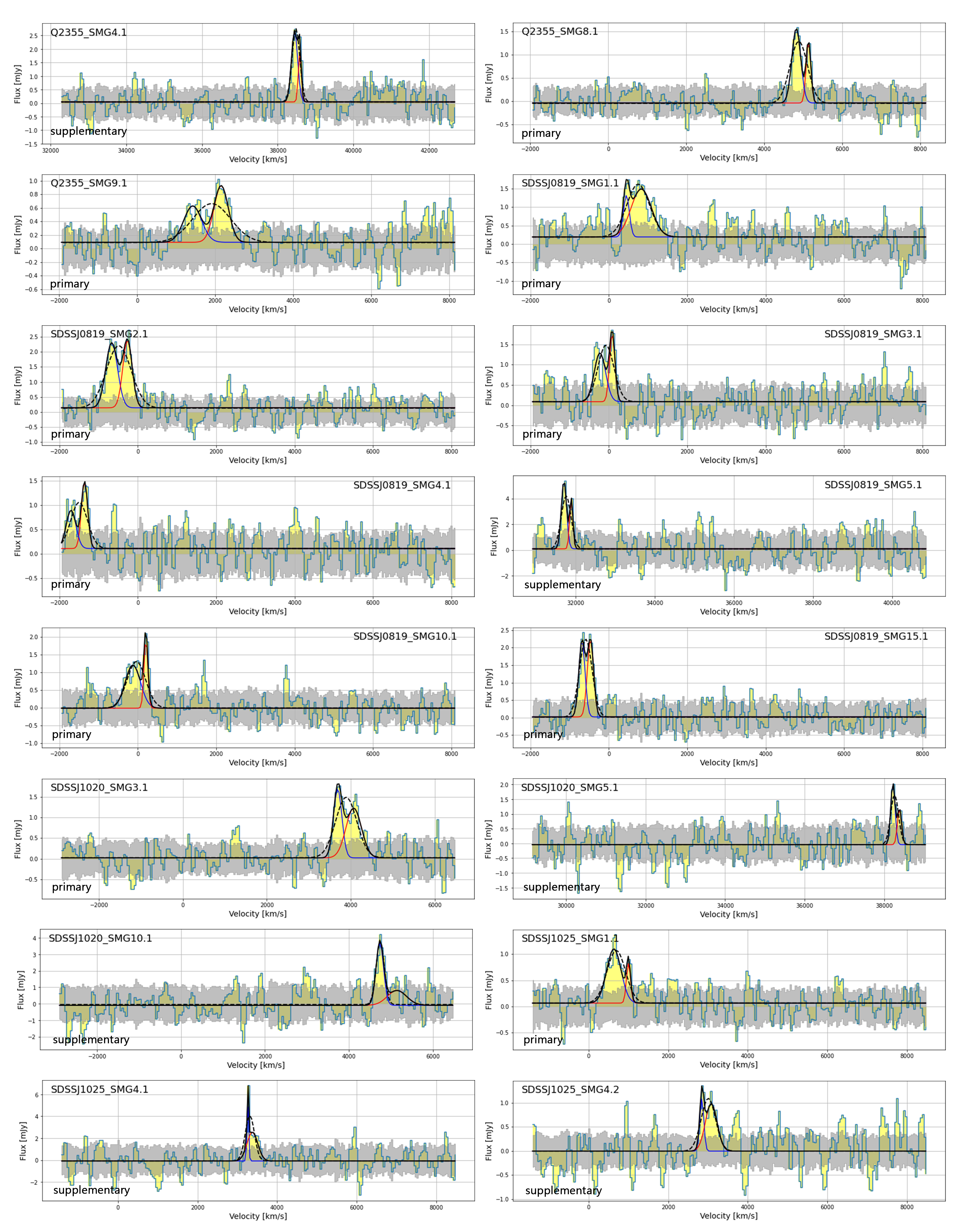}
\end{figure*}

\begin{figure*}[h]
\centering
\includegraphics[width=18cm]{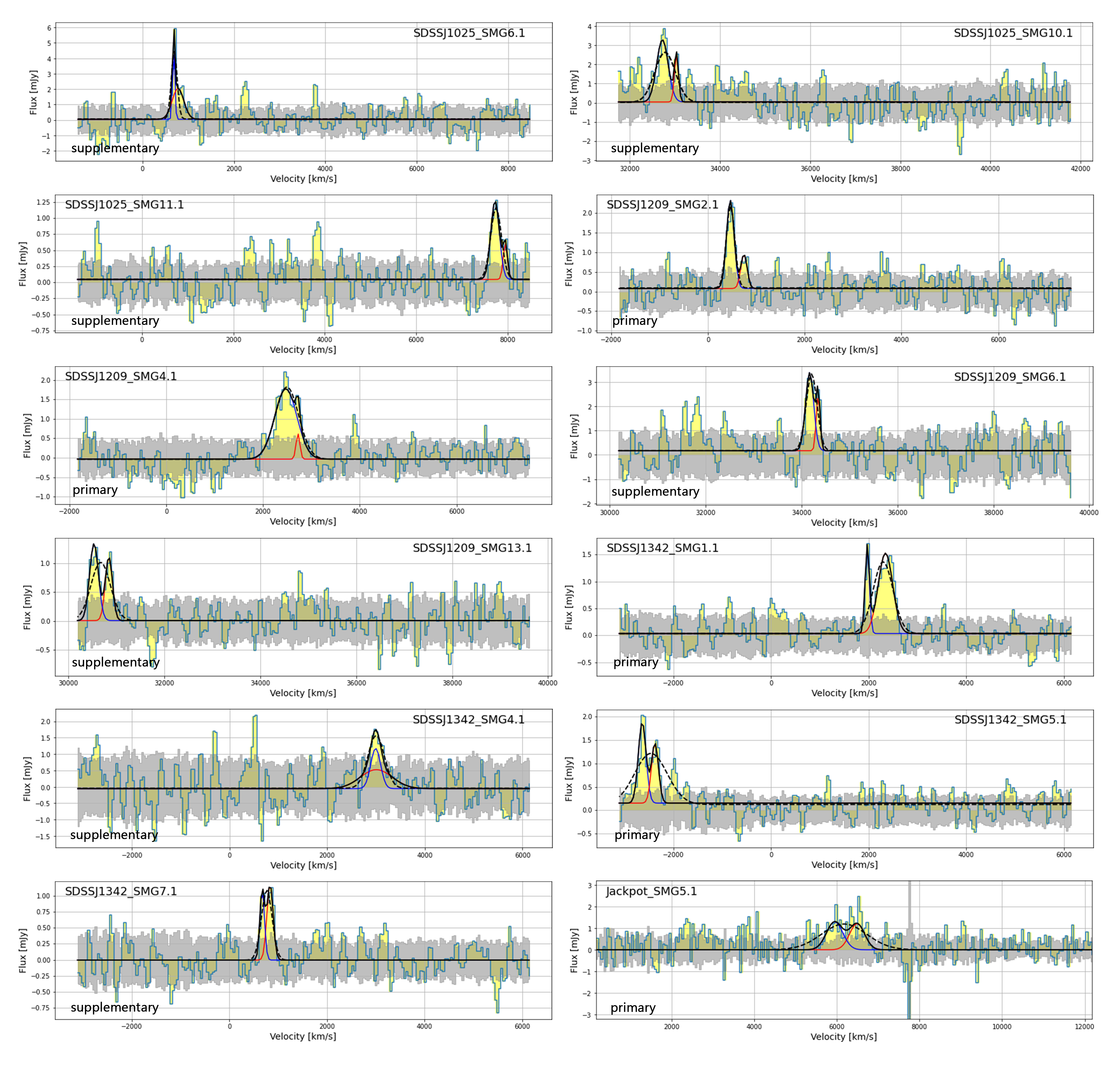}
\caption{CO spectra of all the 28 primary and supplementary detections. The plotting style follows that of the lower panels of Figure~\ref{fig:fig.1b}.
\label{fig:spec2}}
\end{figure*}

\section{CO Luminosity Function} \label{apx:d}

\begin{figure*}[h]
\centering
\includegraphics[width=\textwidth]{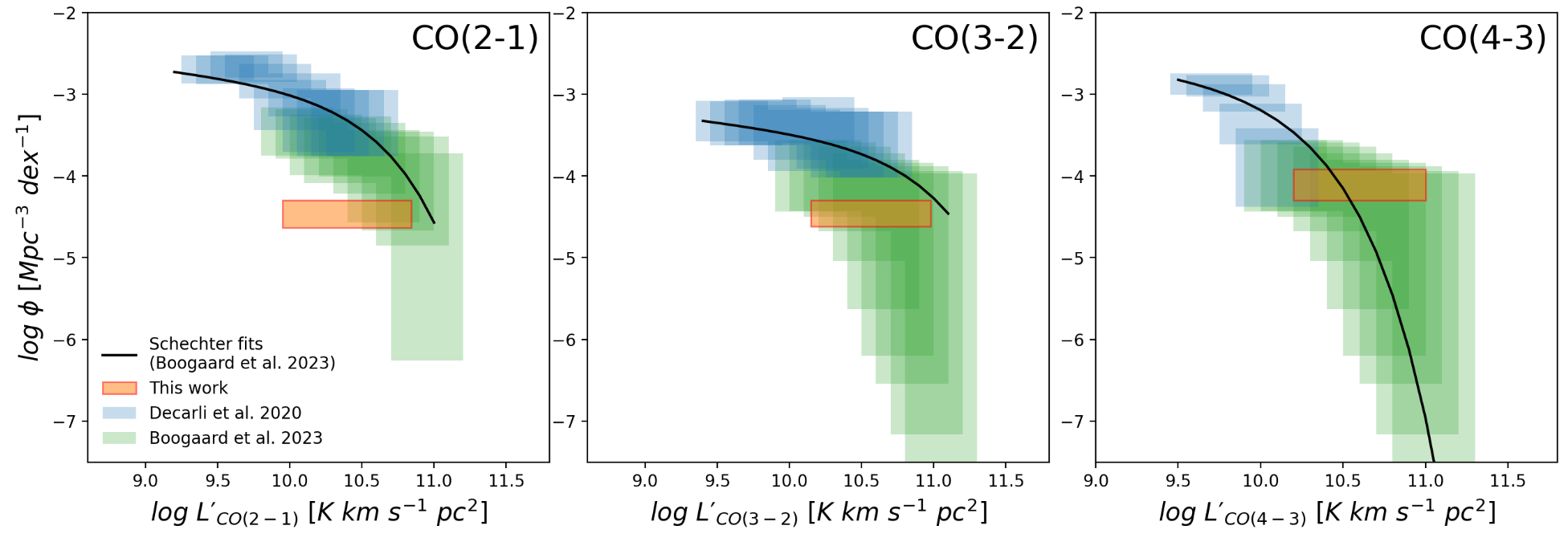}
\caption{CO luminosity functions in various transitions. 
Black curves show the fitting results of Schechter functions reported by \citealp{2023ApJ...945..111B}, while the blue and green blocks show the results from the ASPECS (\citealp{2020ApJ...902..110D}) and HDFN (\citealp{2023ApJ...945..111B}) survey, respectively. Orange blocks show our results for the supplementary sources, which are consistent with those reported in the field, confirming that the supplementary sources are likely field line emitters.
\label{fig:COLF}}
\end{figure*}

In the supplementary sample, there are 7 sources whose velocities are far from the center QSOs and cannot be considered as members of the structure (velocity excess $\pm$ 7000 km/s).
With the aim to explain the existence of these 7 sources and examine their nature, 
we derive their CO luminosity function 
to check whether it is 
consistent with 
that for field SMGs.
First, our boundaries of number density are calculated based on the number of sources and consider their Poisson error, resulting in a range within 1-$\sigma$ uncertainty. The boundaries of CO luminosity are simply calculated using the range between the maximum and minimum CO luminosity among the 7 detections corrected for completeness, and divided by the total survey volume. However, the probabilities of these lines originating from different CO transitions need to be accounted for. 
Considering the fix frequency range of the spectral coverage, a detection will be located at a different redshift when we identify the lines as different CO transitions.
Thus, based on the redshift-dependent probability distribution of field SMGs reported by \citet{2016ApJ...820...82C}, 
we derived the probabilities of these lines being CO(2-1), CO(3-2), CO(4-3), CO(5-4) to be 24.0\%, 57.3\%, 16.6\%, 2.2\%, respectively. We derive the luminosity functions of each transition using these corresponding probabilities. Since the probability of CO(5-4) is very low, we do not report its luminosity function. Subsequently, we plot the results in \autoref{fig:COLF} and we find that they are widely consistent with previous surveys in the field, e.g., ASPECS \citep{2020ApJ...902..110D} and HDFN \citep{2023ApJ...945..111B}. 
Therefore, our analyses suggest that our supplementary sample sources are drawn from the field. We report the expected CO luminosity in different transitions for the supplementary sources in \autoref{tab:diffj}.

\begin{table}[h]
    \centering
    \caption{CO luminosity in diff-$j$ of the supplementary sources}
    \setlength{\tabcolsep}{5mm}{
    \begin{tabular}{lcccc}
    \toprule
        ID &  Completeness& $L'_{\rm CO(2-1)}$ & $L'_{\rm CO(3-2)}$ & $L'_{\rm CO(4-3)}$ \\
          &  &[$10^{10} K km s^{-1} pc^{2}$]& [$10^{10} K km s^{-1} pc^{2}$] & [$10^{10} K km s^{-1} pc^{2}$]\\ \hline
        Q2355\_SMG4.1 & 0.94 & 4.4 ± 1.6 &  5.7 ± 2.0 &  5.8 ± 2.1\\
        SDSSJ0819\_SMG5.1 & 1.00 & 2.6 ± 3.1 &  3.5 ± 4.2 &  3.7 ± 4.5 \\
        SDSSJ1020\_SMG5.1 & 0.86 &2.4 ± 1.1 &  3.2 ± 1.4 &  3.4 ± 1.5 \\
        SDSSJ1025\_SMG10.1 & 0.98 & 7.0 ± 1.8 & 9.5 ± 2.5 &  10.0 ± 2.6 \\
        SDSSJ1025\_SMG11.1 & 0.30 &1.3 ± 0.4 &  2.0 ± 0.6  &  2.2 ± 0.6 \\
        SDSSJ1209\_SMG6.1 & 0.94 &4.4 ± 1.2  &  6.0 ± 1.7 &  6.4 ± 1.8 \\
        SDSSJ1209\_SMG13.1 & 0.96 &2.3 ± 0.5 &  3.3 ± 0.7  &  3.5 ± 0.7 \\
        \hline\hline
    \end{tabular}}
    \label{tab:diffj}
\end{table}

 \section{SED Fitting Samples}\label{apx:f}
\begin{figure*}[h]
 \centering
 \includegraphics[width=\textwidth]{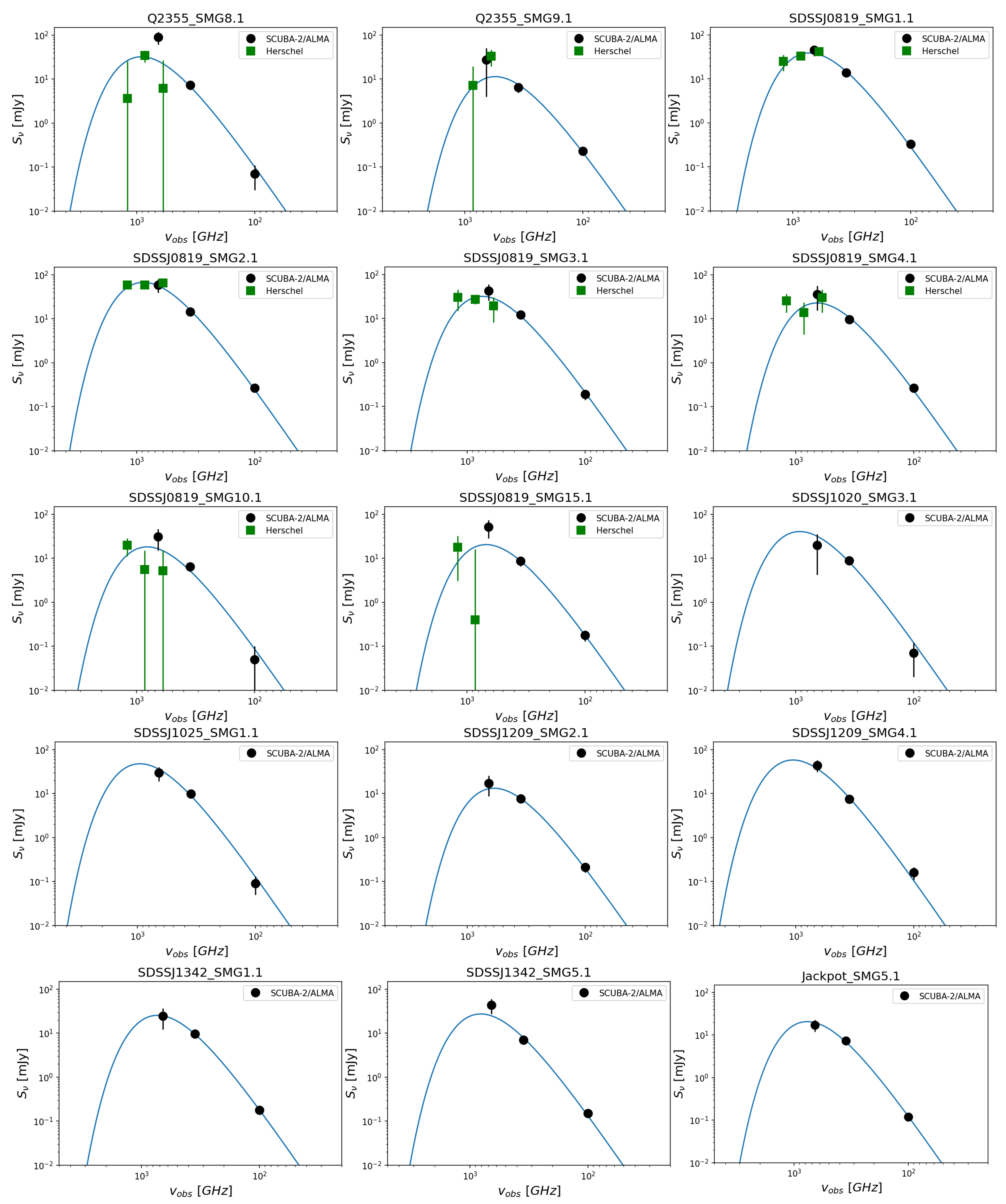}
 \caption{Modified black body (MBB) fitting results for primary sources. The blue line shows the best fitting result. Black points represent the flux density for ALMA and JCMT/SCUBA-2 (from left to right: 450\,$\mu m$, 850\,$\mu m$ (SCUBA-2), 3\,mm (ALMA)), while green points represent those for Herschel (from left to right: 250\,$\mu m$, 350\,$\mu m$, 500\,$\mu m$).
 \label{fig:sedall}}
 \end{figure*}

 We list the results for modified black body (MBB) fitting for primary sources (\autoref{fig:sedall}).

\bibliography{sample631}
\bibliographystyle{aasjournal}


\end{CJK}
\end{document}